\documentclass[10pt]{article}
\usepackage[margin=1in]{geometry}
\usepackage{amsfonts,amsmath,amssymb}
\usepackage[none]{hyphenat}
\usepackage{fancyhdr}
\usepackage{graphicx}
\usepackage{float}
\usepackage[nottoc,notlot,notlof]{tocbibind}
\usepackage{bigints}

\usepackage{amssymb,amsmath,tikz}
\usepackage{pgfplots} 
\usepackage{amsthm}
\usepackage{mathrsfs}
\usepackage{pifont}
\usepackage{slashed}
\usepackage{mathtools}
\usepackage{cancel}

\usepackage[utf8]{inputenc}

\showoutput
\renewcommand{\qedsymbol}{$\blacksquare$}

\theoremstyle{plain}

\newtheorem*{twr*}{Theorem}
\newtheorem*{lem*}{Lemma}
\newtheorem{twr}{Theorem}
\newtheorem{lem}{Lemma}

\newtheorem*{defin*}{Definition}

\newtheorem*{rem*}{Remark}

\newtheorem{cor*}{Corollary}

\newtheorem*{notn*}{Notation}
\newtheorem*{wiener-ito*}{Wiener-It\^o-Segal Decomposition}
\newtheorem*{prop*}{Proposition}

\DeclareMathAlphabet{\mathpzc}{OT1}{pzc}{m}{it}

\begin{document}

\begin{titlepage}
\begin{center}
\vspace*{1cm}
\large{\textbf{FEYNMAN INTEGRAL IN QFT AND WHITE NOISE
ON A COMPACTIFIED VERSION OF A SPACE-TIME WITH 
THE LIE GROUP STRUCTURE
}}\\
\vspace*{2cm}
\small{JAROS{\L}AW WAWRZYCKI}\\[1mm]
\tiny{Bogoliubov Labolatory of Theoretical Physics}
\\
\tiny{Joint Institute of Nuclear Research, 141980 Dubna, Russia}
\\
\vspace*{1cm}
\tiny{\today}\\
\vfill
\begin{abstract}
We present a rigorous construction of the Feynman integral 
on the compactified Einstein Universe (EU) using white noise
calculus. Presented construction of the functional averaging
may also be thought of as a solution to the problem posed by Bogoliubov and Shirkov
in \S 43 of their book \emph{Introduction to the Theory of Quantized
Fields}, Wiley 1980. 
\end{abstract}
\vspace*{0.5cm}
\tiny{{\bf Keywords}: functional averaging, Feynman integral, \\
white noise, Hida operators}
\end{center}
\vfill
\end{titlepage}



\section{Introduction}

The white noise calculus and Hida operators are essential for the results we have achieved 
for perturbative QFT on the Minkowski space-time \cite{wawrzyckiInfinite} -- \cite{wawrzyckiModPhys}, 
where we have reduced the freedom in the choice of renormalization and
obtained nontrivial restrictions on the particle masses. 
The analog results for perturbative
QFT on EU could have been achieved without the use of white noise analysis, compare \cite{SegalZhouQED}.
But in going beyond the perturbative methods to computation of the full Green functions
\[
\Delta(x,y) = i {\textstyle\frac{\left\langle T \boldsymbol{\varphi}(x)\boldsymbol{\varphi}(y) S \right\rangle_{{}_{0}}}{\langle S \rangle_{{}_{0}}}},
\,\,\,\,\,
\textrm{$\langle \cdot \rangle_{{}_{0}}$ -- vacuum average},
\] 
which are based on the functional averaging, we found the white noise Hida test space and the Hida 
generalized functions indispensable also on EU.

We present here a rigorous construction of the integral averaging, based on the method outlined in
\cite{Bogoliubov_Shirkov}, \S\S 43.1 -- 43.4, and performed within the momentum representation.
The Einstein Universe (EU) is naturally the Lie group $\mathbb{R} \times SU(2,\mathbb{C})$, 
and the free fields have a common period on EU,  \cite{PaneitzSegalI}
-\cite{PaneitzSegalIII}. We consider the compactified EU $\simeq G = [\mathbb{R} \, \textrm{mod} \, 4\pi] \times SU(2,\mathbb{C})$
with the period $4\pi$ common for the period of all free fields on EU. 

For the compactified EU $\simeq G = [\mathbb{R} \, \textrm{mod} \, 4\pi] \times SU(2,\mathbb{C})$, 
and periodic $\theta$ (in ret and av parts in chronological products), the Fourier transform has purely discrete 
range, which allows us to avoid passing to the continuum
limit of the original idea of \cite{Bogoliubov_Shirkov}, \S\S 43.1. We construct the averaging
for the compactified EU and then eventually pass to the case in which $\theta$ has larger, 
but always an integral multiple of the period of the  free fields on EU, but arbitrarily
large period. Here we recall the idea of \cite{Bogoliubov_Shirkov}, \S 43,
but applied to the compactified EU, and propose spaces of functions $\nu,\varphi$ 
to ensure the convergence of the method -- the problem posed there. 

For simplicity, we consider after \cite{Bogoliubov_Shirkov} the real (essentially neutral)
scalar field $\boldsymbol{\varphi}(x)$ on $G$, and whenever we are using bold letter, we mean
the free field operator, and not ordinary functions. Ordinary functions are written with non-bold letters. 

We are using the standard Fourier transform $\varphi \mapsto \widetilde{\varphi}$ on $G$
and the ordinary invariant $L^2$-inner product $\langle \cdot, \cdot \rangle$ for (here real) functions on
$G$ with the ordinary invariant measure on $G$, normalized to $4\pi$ (our $G$ is compact), 
with the unitary characters (unitary irreducible
representations) of $G$ equal
\[
G \ni (t,w) \mapsto \widehat{n} \cdot \widehat{l}(t,w) = \widehat{n}(t)\widehat{l}(w) 
\]    
with $\widehat{n}(t) = \exp[i nt/2]$, and with $\widehat{l}$ equal to the standard 
unitary irreducible representation of $SU(2,\mathbb{C})$ with weight $l$. 
$n \in \mathbb{Z}$ and $l$ range over a non-negative integer or half an integer
numbers. 

The practical motivation for our work is provided at the very end.

\section{Fundamental Gelfand triples}\label{GelfandTriples}

We start at once with the definition of the nuclear spaces within which we will be seeking
the proper domains for $\nu,\varphi$. First we use the real space-time Gelfand triple
\[
E = \mathcal{S}_{A}(G;\mathbb{R}) \subset L^2(G;\mathbb{R}) = H \subset  \mathcal{S}_{A}(G;\mathbb{R})^* = E^*
\]  
with the standard operator $A = -4\partial_{t}^2 \otimes 1 - 1 \otimes \Delta_{{}_{SU(2,\mathbb{C})}} +1$ on $L^2(G;\mathbb{R})$,
and its Bose-Fock lifting to the Gelfand triple over the Bose-Fock space $\Gamma(H)$:
\[
(E) = \mathcal{S}_{\Gamma(A)}(E^*;\mathbb{R}) \subset L^2(E^*, d\mu) = \Gamma(H) \subset  \mathcal{S}_{\Gamma(A)}(E^*;\mathbb{R})^* = (E)^*.
\] 
Here $\mu$ is the standard Gaussian probability measure on $E^*$, with the canonical isomorphism
$L^2(E^*, d\mu) = \Gamma(H)$ given by the Wiener-It\^o-Segal chaos decomposition, compare 
\cite{obata-book}, \cite{GelfandIV}. Below, we will be referring also to theorems
involving complexifications $E_{\mathbb{C}}, H_\mathbb{C}, E_\mathbb{C}^*$ of the spaces $E,H,E^*$
and their Fock liftings, but we should emphasize that the real versions of these spaces play a distinguished
role not only for the real field, and the existence of the measure $\mu$ is associated with the reality of the
considered spaces. In what follows, we identify the spaces  $E,H,E^*$ with their Fourier transform
isomorphic images $\widetilde{E},\widetilde{H},\widetilde{E}^*$, and correspondingly
we identify their Fock liftings, with the inner product canonically 
inducing the dual pairings $\langle \nu, \varphi \rangle = \langle \widetilde{\nu}, \widetilde{\varphi}\rangle$, 
given by the Plancherel formula, with the standard operator $\widetilde{A}$ equal to the Fourier transform of $A$.
The dual paring in the Fock lifted Gelfand triple is induced by the inner product $\langle\langle \cdot, \cdot \rangle\rangle$
in $\Gamma(H)$, which in the function representation, via the Wiener-It\^o-Segal isomorphism, is equal to the
$L^2$ norm on  $L^2(E^*, d\mu)$.  In the sequel, we will also use the Hilbert
spaces $E_k$, $k \in \mathbb{Z}$ -- the closures of the domains of $A^k$ in $H$, with respect to the Hilbertian norms 
$| \cdot |_{{}_{k}} \overset{def}{=} |A^k \cdot|_{{}_{0}}$, where $|\cdot |_{{}_{0}}$ is the Hilbert space norm of $H$.
The space $E$ becomes then a projective limit $\cap_{{}_{k}} E_k$ of $E_k$, with its dual equal to the inductive limit 
$E^* = \cup_{{}_{k}} E_k$. We again identify these spaces with their Fourier transform images, $\widetilde{E}_k$,
with the standard operator equal $\widetilde{A}$. We have the natural continuous inclusions 
\[
E \subset \ldots E_{k} \subset \ldots E_{k-1} \subset \ldots \subset E_{1} \subset E_0 = H 
\subset E_{-1} \subset E_{-k +1} \subset \ldots E_{-k} \subset \ldots \subset E^* 
\] 
with $E^*_{k}$ canonically isomorphic to $E_{-k}$ for $k\in \mathbb{Z}$ (in fact, we can use any real numbers for $k$).
We have analogous Hilbert $(E)_k$ spaces and continuous inclusions, and the realizations as the projective and inductive limits
for the Hida nuclear spaces $(E), (E)^*$, 
with the standard operator $A$ replaced with the standard operator $\Gamma(A)$. For a more detailed description of this construction
of Hida test space and its dual, we recommend \cite{obata-book}. In what follows we use the fact, already exploited by us in the construction
of the free fields on EU, that we have at our disposal a natural way of constructing closed subspaces of the Gelfand triple
$E\subset H\subset E^*$, which again compose Gelfand triples naturally embedded in the original Gelfand triple $E\subset H\subset E^*$. 
Indeed, we can do it just by passing to the Fourier images $\widetilde{E}\subset \widetilde{H}\subset \widetilde{E}^*$ 
and by selecting any subset $\mathscr{R}$ of the characters $\widehat{n}\cdot\widehat{l}$, 
and restrict the elements of the spaces to these and only these
elements with the Fourier transform, supported at this subset. The corresponding standard operator becomes equal to 
the restriction of $\widetilde{A}$ to the elements of the initial $\widetilde{H}$, with the support at this fixed subset $\mathscr{R}$.
The corresponding closed subspaces $E_{{}_{\mathscr{R} \,\, k}} \subset E_{{}_{k}}$, 
and their productive and inductive limits, composing the corresponding Gelfand triples 
$E_{{}_{\mathscr{R}}} \subset H_{{}_{\mathscr{R}}} \subset E_{{}_{\mathscr{R}}}^*$, 
will frequently be denoted without the subscript $\mathscr{R}$,
in order to simplify notation, whenever it is clear whether we mean an element $\widetilde{\xi} \in \widetilde{E}_{{}_{k}}$
supported at $\mathscr{R}$, and belonging to $\widetilde{E}_{{}_{\mathscr{R} \,\, k}}$, or generally whenever it is clear that
we mean $\widetilde{\xi}$ supported at $\mathscr{R}$. In particular, the Gelfand triples
corresponding to the real field with the positive energy orbit $\mathscr{O}^+$, are constructed from the Gelfand
triple $\widetilde{E}\subset \widetilde{H}\subset \widetilde{E}^*$, just by restriction of the Fourier transform components 
to the subset $\mathscr{R} = \mathscr{O}^+$.
 
Note here, that also each  element of $\widetilde{E}^*$ can be regarded as a Fourier series, but the condition
$\widetilde{\varphi} \in \widetilde{E}^*$ is equivalent with the existence of $k$, such that  $\widetilde{\varphi} \in \widetilde{E}_{-k}$,
which, in turn, is equivalent to
\[
\sum (2l+1) \textstyle{\frac{1}{(n^2 + 1 + l(l+1))^{2k}}}
\widetilde{\varphi}(\widehat{n}\cdot\widehat{l})_{{}_{i \, j}}\overline{\widetilde{\varphi}(\widehat{n}\cdot\widehat{l})_{{}_{i \, j}}}
< +\infty, 
\]
with the summation range $n\in \mathbb{Z}, l \in \mathbb{N}, -l \leq i,j \leq l$ for the initial Gelfand triple
$\widetilde{E}\subset \widetilde{H}\subset \widetilde{E}^*$ or, respectively, with the
summation range $\widehat{n}\cdot\widehat{l} \in \mathscr{R}$, $ -l \leq i,j \leq l$, for 
$\widetilde{\varphi} \in E_{{}_{\mathscr{R}}}^*$ defined by the restriction of the support to  $\mathscr{R}$. Below, we consider real
$\nu, \varphi$, so that 
\[
\overline{\widetilde{\varphi}(\widehat{n}\cdot\widehat{l})_{{}_{i \, j}}} = \widetilde{\varphi}(\widehat{-n}\cdot\widehat{l})_{{}_{-i \, -j}},
\,\,\,
\overline{\widetilde{\nu}(\widehat{n}\cdot\widehat{l})_{{}_{i \, j}}} = \widetilde{\nu}(\widehat{-n}\cdot\widehat{l})_{{}_{-i \, -j}} 
\]
and for real $\varphi$
\begin{multline*}
\langle \widetilde{\varphi}, \widetilde{\varphi} \rangle
=
\sum (2l+1)
\widetilde{\varphi}(\widehat{n}\cdot\widehat{l})_{{}_{i \, j}}\overline{\widetilde{\varphi}(\widehat{n}\cdot\widehat{l})_{{}_{i \, j}}}
\\
= 
\sum (2l+1)
\widetilde{\varphi}(\widehat{n}\cdot\widehat{l})_{{}_{i \, j}} \widetilde{\varphi}(\widehat{-n}\cdot\widehat{l})_{{}_{-i \, -j}}
\\
=
\sum (2l+1) \left[
x_{{}_{\widehat{n}, \widehat{l} \, i \, j}}^2  + y_{{}_{\widehat{n}, \widehat{l} \, i \, j}}^2
\right]
\end{multline*}
if 
\begin{multline*}
\widetilde{\varphi}(\widehat{n}\cdot\widehat{l})_{{}_{i \, j}} =
x_{{}_{\widehat{n}, \widehat{l}, \, i \, j}} + i y_{{}_{\widehat{n}, \widehat{l}, \, i \, j}}^2,
\\
\textrm{Re}  \, \left[ \widetilde{\varphi}(\widehat{n}\cdot\widehat{l})_{{}_{i \, j}} \right] = 
x_{{}_{\widehat{n}, \widehat{l}, \, i \, j}},
\,\,\,
\textrm{Im}  \, \left[ \widetilde{\varphi}(\widehat{n}\cdot\widehat{l})_{{}_{i \, j}} \right] = 
y_{{}_{\widehat{n}, \widehat{l}, \, i \, j}},
\end{multline*}
because the real part is symmetric, and the imaginary part is skew-symmetric:
\[
x_{{}_{\widehat{-n}, \widehat{l}, \, -i \, -j}} = x_{{}_{\widehat{n}, \widehat{l} \, i \, j}},
\,\,\,
y_{{}_{\widehat{-n}, \widehat{l}, \, -i \, -j}} = - y_{{}_{\widehat{n}, \widehat{l}, \, i \, j}}.
\]

We will do computations in the momentum representation (after Fourier transform). Recall that the free field has the 
form of the Fourier transform, which, in case of compactified EU $=G$ (and also for ordinary non-compactified EU), 
has the form of the Fourier series with well-defined Fourier-operator coefficients 
$\widetilde{\boldsymbol{\varphi}}(\widehat{n}\cdot\widehat{l})_{{}_{i \,\, j}}$.
This is the consequence of the $\zeta^4$invariance \cite{PaneitzSegalI}-\cite{PaneitzSegalIII}.   
This series is moreover finite, if the field $\boldsymbol{\varphi}$ is massive. For the real scalar field, we have
\[
\widetilde{\boldsymbol{\varphi}}(\widehat{n}\cdot\widehat{l})_{{}_{i \,\, j}} =
\begin{cases}
{\textstyle\frac{\sqrt{4\pi}}{2l+1}}\sum\limits_{s=1}^{(2l+1)^2}\left[
u_{{}_{s}}(\widehat{n}\cdot\widehat{l})_{{}_{i \,\, j}} a_{{}_{s}}(\widehat{n}\cdot\widehat{l})
+
\overline{u_{{}_{s}}(\widehat{n}\cdot\widehat{l})_{{}_{-j \,\, -j}}} a_{{}_{s}}(\widehat{n}\cdot\widehat{l})^+\right], & 
\textrm{if} \,\,\, 
\widehat{n}^+\cdot\widehat{l} \in \mathscr{O}^+, \\
0, &  \textrm{if} \,\,\, 
\widehat{n}^+\cdot\widehat{l} \notin \mathscr{O}^+,
\end{cases}
\]
where $\mathscr{O}^+$ is the positive energy orbit
corresponding to the real scalar field ($\mathscr{O}^+, \mathscr{O}^-$ are finite sets if the field is massive, compare \cite{SegalZhouQED}).
The orbit is determined by the invariant equation respected by the free field and the finite range
matrices $u_{{}_{s}}(\widehat{n}\cdot\widehat{l})$ are the Fourier transforms of fundamental solutions, concetrated on the single point character
set $\{ \widehat{n}\cdot\widehat{l}\}$ (non-scalar fields will have $u$ with one more index $a$ counting the number of components
of the field $\boldsymbol{\varphi}$). Aso the finite range of the discrete index $s$ depends 
on the actual character $\widehat{n}\cdot\widehat{l}$ and on the kind of field. For the scalar field $s$ ranges among all positive integers
from $1$ to $(2l+1)^2$, and below in the sums we will not write it explicitly in order to simplify notation.   
$a_{{}_{s}}(\widehat{n}\cdot\widehat{l}), a_{{}_{s}}(\widehat{n}\cdot\widehat{l})^+$ are 
the Hida annihilation-creation operators, which for real scalar field in addition fulfil
\[
a_{{}_{s}}(\widehat{n}\cdot\widehat{l}) = a_{{}_{s}}(\widehat{-n}\cdot\widehat{l}),
\,\,\,
\left[a_{{}_{s}}(\widehat{n}\cdot\widehat{l}), a_{{}_{s'}}(\widehat{n'}\cdot\widehat{l'})^+ \right] = 
\delta_{{}_{s \, s'}} \delta_{{}_{n \, n'}} \delta_{{}_{l \, l'}}.
\]
For non-scalar fields, we have an analogous generalization 
with various roles of the positive and negative energy parts of the corresponding orbit, depending on if the field is neutral (real)
or charged (complex).  The method of computation of the matrices $u$ is based on the decomposition of
the representation generated by the transformation rule, and is based on the Fourier transform decomposition. 
In the computation of $u$ we include only these
$u$ which lead to the field invariant under $\zeta$ or $\zeta^4$, where $\zeta(t,w)= (t+\pi, \textrm{Antipode of} \, w)$.
For the real scalar field, $\zeta\boldsymbol{\varphi} = \boldsymbol{\varphi}$. 
$u$ can also be computed by comparison with the classification
of the fields on EU given in \cite{PaneitzSegalI}-\cite{PaneitzSegalIII}. 

For the compactified EU, with the $\theta$-function of the same period as the common 
period of all free fields on EU, also such periodic $\theta$ can be Fourier transformed and decomposed into the same complete
system of characters $\widehat{n}\cdot\widehat{l}$, as the free fields themselves.
In particular, using the Fourier transform isomorphism, 
the vacuum average of the chronological product (chronological pairing) 
$\left\langle T(\boldsymbol{\varphi}(x_1)\boldsymbol{\varphi}(x_2))\right\rangle_{{}_{0}} $ can be written
in the momentum picture in the form
\begin{equation}\label{chronologicalProd}
\left\langle T\left(\widetilde{\boldsymbol{\varphi}}(\widehat{n_1}\cdot\widehat{l_1})_{{}_{i_1 \,\, j_1}} 
\widetilde{\boldsymbol{\varphi}}(\widehat{n_2}\cdot\widehat{l_2})_{{}_{i_2 \,\, j_2}}\right)\right\rangle_{{}_{0}} 
=
-i{\textstyle\frac{\sqrt{4\pi}}{2l_1+1}}
\widetilde{\Delta_{c}}(\widehat{n_1}\cdot\widehat{l_1})_{{}_{j_1 \,\, j_1}}
\delta_{{}_{n_1 \, -n_2}} \delta_{{}_{i_1 \, -i_2}} \delta_{{}_{j_1 \, -j_2}} \delta_{{}_{l_1 \, l_2}},
\end{equation}
where
\begin{multline*}
{\textstyle\frac{1}{\sqrt{4\pi}^2}} \int\limits_{G\times G} dt_1dw_1dt_2dw_2 \Delta_c(t_1-t_2,w_1w_2^{-1})  
\overline{\widehat{n_1}(t_1)} \, \overline{\widehat{l_1}(w_1)_{{}_{j_1 \, i_1}}} \,
\overline{\widehat{n_2}(t_2)} \, \overline{\widehat{l_2}(w_2)_{{}_{j_2 \, i_2}}} 
\\
=
{\textstyle\frac{\sqrt{4\pi}}{2l_1+1}}
\widetilde{\Delta_{c}}(\widehat{n_1}\cdot\widehat{l_1})_{{}_{j_1 \,\, j_1}}
\delta_{{}_{n_1 \, -n_2}} \delta_{{}_{i_1 \, -i_2}} \delta_{{}_{j_1 \, -j_2}} \delta_{{}_{l_1 \, l_2}}   
\end{multline*}
is the Fourier transform of the function $(x_1,x_2) \rightarrow \Delta_c(x_1x_2^{-1})$, treated as the function of two independent space-time 
variables, $x_1=(t_1,w_1),x_2=(t_2,w_2)$, and such that 
\[
\left\langle T(\boldsymbol{\varphi}(x_1)\boldsymbol{\varphi}(x_2))\right\rangle_{{}_{0}}  = D_c(x_1x_2^{-1}) = -i\Delta_c(x_1x_2^{-1})
\]
is the chronological pairing of the free real field,
and where
\[
\widetilde{\Delta_{c}}(\widehat{n}\cdot\widehat{l})_{{}_{j \,\, i}}
=
{\textstyle\frac{1}{\sqrt{4\pi}}} \int\limits_{G} dtdw \,\, \Delta_c(t,w)  
\overline{\widehat{n}(t)} \, \overline{\widehat{l}(w)_{{}_{i \, j}}}
\] 
is the Fourier transform of $\Delta_c$, treated as a function of one space-time variable $x=(t,w)$. Recall
that for massive field on EU, the causal distribution $D_c$ is equal to ordinary function
and its Fourier transform represents an $L^2$-summable series. We use the reality and parity of $\Delta_c$,
which together imply
\[
\widetilde{\Delta_{c}}(\widehat{n}\cdot\widehat{l})_{{}_{j \,\, i}} = \widetilde{\Delta_{c}}(\widehat{-n}\cdot\widehat{l})_{{}_{-i \,\, -j}},
\,\,\,\,
\widetilde{\Delta_{c}}(\widehat{n}\cdot\widehat{l})_{{}_{j \,\, j}} = \widetilde{\Delta_{c}}(\widehat{-n}\cdot\widehat{l})_{{}_{-j \,\, -j}}.
= \overline{\widetilde{\Delta_{c}}(\widehat{n}\cdot\widehat{l})_{{}_{j \,\, j}} }
\]

\section{The functional averaging due to Bogoliubov}

Let us first briefly give the general idea of \cite{Bogoliubov_Shirkov}, \S 43, and only then will we pass to a more detailed
discussion. It is suggested there to compute first the vacuum averaging of the chronological product of the simple 
functional $e^{i\langle \nu,\boldsymbol{\varphi}\rangle}$ 
$=e^{i\langle \widetilde{\nu},\widetilde{\boldsymbol{\varphi}}\rangle}$ -- the quantum counterpart of the
character functional $F(\widetilde{\varphi}) = e^{i\langle \nu,\varphi\rangle} = e^{i\langle \widetilde{\nu},\widetilde{\varphi}\rangle}$ 
of the nuclear group $E$, with addition of vectors in $E$ as the group action,
and with $\varphi$ presumably coming from a closed subspace of $E^*$, and $\nu$ representing a functional from $E^*$,
which can sensibly be restricted to this subspace. The closer identification of these subspaces will come only 
after a more detailed analysis of the method of \cite{Bogoliubov_Shirkov}, \S 43. The determination of these subspaces 
will be reduced to the determination of the proper support $\mathscr{R}$, mentioned above. In the case of the character
functional, the computation of  
\begin{equation}\label{I(nu)}
\mathcal{T}\mu_T\left(\widetilde{\nu}\right)  = \left\langle T\left( e^{i\langle \nu,\boldsymbol{\varphi}\rangle}\right)\right\rangle_{{}_{0}} 
= \left\langle T\left( e^{i\langle \widetilde{\nu},\widetilde{\boldsymbol{\varphi}}\rangle}\right)\right\rangle_{{}_{0}}
\end{equation}
in terms of a  ``functional averaging'' can accordingly be simplified, \cite{Bogoliubov_Shirkov}, \S 43.1-43.2,
when passing to the momentum representation:
\begin{equation}\label{BSstep1}
\mathcal{T}\mu_T\left(\widetilde{\nu}\right) = \int\limits_{E^*} e^{i\langle \widetilde{\nu},\widetilde{\varphi}\rangle} d\mu_T(\widetilde{\varphi}),
\end{equation}
in which we have an operation closely related to integration over $E^*$ (we will identify it later).
The next essential point of the idea of Bogoliubov and Shirkov, consists in passing from the 
simple character functional to more involved functionals $F(\widetilde{\varphi})$ which, hopefully, can be
subjected to the $d\mu_T$-``integration''. First of all, they ask for the functionals $F$ (in general, nonlinear)
which are equal to Fourier transforms $\mathcal{T} \Lambda$ of ``something'' $\Lambda$ (let say here, measures, or more generally,
Hida infinite dimensional distributions $\Lambda$ lying in $(E)^* = \mathcal{S}_{\Gamma(A)}(E^*;\mathbb{R})^*$):
\begin{equation}\label{BSstep2}
F(\widetilde{\varphi}) = \int\limits_{E^*} e^{i\langle\widetilde{\nu},\widetilde{\varphi}\rangle} \Lambda(\widetilde{\nu})
d\mu(\widetilde{\nu}),
\end{equation}
here with the standard Gaussian probability measure $\mu$ on $E^*$. 
Having a functional $F$ as the Fourier transform image of a distribution or a measure, we define
after \cite{Bogoliubov_Shirkov}, \S 43.1-43.2: 
\begin{multline}\label{BSstep3}
\left\langle T\left(F(\boldsymbol{\varphi})  \right)\right\rangle_{{}_{0}} 
= \int\limits_{E^*} \left\langle T\left( e^{i\langle \nu,\boldsymbol{\varphi}\rangle}\right)\right\rangle_{{}_{0}}
 \Lambda(\widetilde{\nu})
d\mu(\widetilde{\nu})
\\
=
\int\limits_{E^*} \int\limits_{E^*} e^{i\langle \widetilde{\nu},\widetilde{\varphi}\rangle} d\mu_T(\widetilde{\varphi}) 
\Lambda(\widetilde{\nu})
d\mu(\widetilde{\nu})
\\
=
\int\limits_{E^*} \int\limits_{E^*} e^{i\langle \widetilde{\nu},\widetilde{\varphi}\rangle} \Lambda(\widetilde{\nu})
d\mu(\widetilde{\nu})
d\mu_T(\widetilde{\varphi}) 
= 
\int\limits_{E^*} F(\widetilde{\varphi})
d\mu_T(\widetilde{\varphi}).
\end{multline}
Here for the functionals $F,G$ for which $F(\varphi) = G(\widetilde{\varphi})$, we use the same 
symbol $F$, i.e. we denote $G$ also by $F$, in order to simplify notation, 
and use $\mathcal{T}$ for the infinite dimensional Fourier transform, which we define below.

Let us now take a closer look at each of the three steps (\ref{BSstep1})-(\ref{BSstep3}), and try to choose
the spaces of allowed $\nu$ and $\varphi$ more closely, which make the definition of $d\mu_T$ rigorous,
and will give meaning and secure the convergence of limit transitions used in the computations.

\section{Convergence of $T\left( e^{i\langle \nu,\boldsymbol{\varphi}\rangle}\right)$}

The computation of (\ref{BSstep1}) is proposed to be performed in two stages in \cite{Bogoliubov_Shirkov}. In the first
stage, we compute explicitly
\[
T\left( e^{i\langle \nu,\boldsymbol{\varphi}\rangle}\right), 
\]
as an operator series, in terms of the components of $\widetilde{\nu}$ and the operator components of
$\widetilde{\boldsymbol{\varphi}}$ and then compute
\[
\left\langle T\left( e^{i\langle \nu,\boldsymbol{\varphi}\rangle}\right)\right\rangle_{{}_{0}} 
\]
in terms of an exponent of a series involving $\widetilde{\nu}, \widetilde{\varphi}$, regarded as 
elements of $E^*$, without any operators. 

First we compute
\[
T\left(F(\boldsymbol{\varphi})  \right)
\]
for the functionals of the following monomial form
\[
\widetilde{\boldsymbol{\varphi}} \mapsto \langle \widetilde{\nu},\widetilde{\boldsymbol{\varphi}}\rangle^m, 
\,\,\, 
\widetilde{\boldsymbol{\varphi}} \mapsto \langle \widetilde{\nu_1},\widetilde{\boldsymbol{\varphi}}\rangle \ldots 
\langle \widetilde{\nu_m},\widetilde{\boldsymbol{\varphi}}\rangle
\]  
which correspond to the classical functionals $F$, which are symmetric monomials
\begin{multline*}
\widetilde{E}^* \ni \widetilde{\varphi} \mapsto \langle \widetilde{\nu},\widetilde{\varphi}\rangle^m 
= \langle \widetilde{\nu}^{\otimes \, m}, \widetilde{\varphi}^{\otimes \, m} \rangle,
\\
\widetilde{E}^* \ni \widetilde{\varphi} \mapsto \langle \widetilde{\nu_1},\widetilde{\varphi}\rangle \ldots \langle \widetilde{\nu_m},\widetilde{\varphi}\rangle
= \left\langle \widetilde{\nu}_1 \otimes \ldots \otimes \widetilde{\nu_m}, \widetilde{\varphi}^{\widehat{\otimes} m} \right\rangle,
\,\,\, \widetilde{\nu}, \widetilde{\nu_i} \in \widetilde{E}
\end{multline*}
on $\widetilde{E}^*$ or, equivalently, on $E^*$. Note that the role of $\nu$ and $\phi$ is symmetric, 
as we can exchange here $E$ and $E^*$ with each other, or if $\nu, \nu_i \in E_k$, then $\varphi \in E_{-k}$, 
in order to make the pairings well-defined. We will specify 
the domains of admissible $\nu, \nu_i$ and $\varphi$ soon. $\widehat{\otimes}$ denotes symmetrized tensor product 
(which for nuclear spaces is uniquely defined, as the projective and injective tensor products coincide in this case). 
When we use the tensor product of Hilbert spaces, we mean the Hilbert-space tensor product.

For example, by the Wick theorem and the formula (\ref{chronologicalProd}) for the chronological pairing,
we obtain
\begin{multline*}
T\left( \langle \widetilde{\nu},\widetilde{\boldsymbol{\varphi}}\rangle^2\right) 
= T\left( \sum\limits_{\widehat{l_1},\widehat{n_1}} (2l_1+1)
\textrm{Tr} \, \left[\widetilde{\boldsymbol{\varphi}}(\widehat{n_1}\cdot\widehat{l})
\widetilde{\nu}(\widehat{n_1}\cdot\widehat{l_1})\right] 
\sum\limits_{\widehat{l_2},\widehat{n_2}} (2l_2+1)\textrm{Tr} \, 
\left[\widetilde{\boldsymbol{\varphi}}(\widehat{n_2}\cdot\widehat{l_2})
\widetilde{\nu}(\widehat{n_1}\cdot\widehat{l_2})\right]
\right) 
\\
= \,\,\,\,
{:}
\left( \sum\limits_{\widehat{l_1},\widehat{n_1}} (2l_1+1)
\textrm{Tr} \, \left[\widetilde{\boldsymbol{\varphi}}(\widehat{n_1}\cdot\widehat{l_1})
\widetilde{\nu}(\widehat{n_1}\cdot\widehat{l_1})\right] 
\sum\limits_{\widehat{l_2},\widehat{n_2}} (2l_2+1)
\textrm{Tr} \, \left[\widetilde{\boldsymbol{\varphi}}(\widehat{n_2}\cdot\widehat{l_2})
\widetilde{\nu}(\widehat{n_2}\cdot\widehat{l_2})\right]
\right) 
{:}
\\
-i \sum\limits_{\widehat{l},\widehat{n}, i,j} (2l+1) 
\widetilde{\nu}(\widehat{n}\cdot\widehat{l})_{{}_{j \, i}} \,\,
\sqrt{4\pi} \widetilde{\Delta_{c}}(\widehat{n}\cdot\widehat{l})_{{}_{jj}}
\,\,
\widetilde{\nu}(\widehat{-n}\cdot\widehat{l})_{{}_{-j \, -i}},
\end{multline*}
\begin{multline*}
=
{:}
\left( \sum\limits_{\widehat{l_1}\cdot\widehat{n_1} \in \mathscr{O}^+} (2l_1+1)
\textrm{Tr} \, \left[\widetilde{\boldsymbol{\varphi}}(\widehat{n_1}\cdot\widehat{l_1})
\widetilde{\nu}(\widehat{n_1}\cdot\widehat{l_1})\right] 
\sum\limits_{\widehat{l_2}\cdot\widehat{n_2} \in \mathscr{O}^+} (2l_2+1)
\textrm{Tr} \, \left[\widetilde{\boldsymbol{\varphi}}(\widehat{n_2}\cdot\widehat{l_2})
\widetilde{\nu}(\widehat{n_2}\cdot\widehat{l_2})\right]
\right) 
{:}
\\
-i \left\langle \widetilde{\nu}, \widetilde{T_c} \widetilde{\nu}^*\right\rangle \,\,\,
= \,\,\,\, {:} \langle \widetilde{\nu},\widetilde{\boldsymbol{\varphi}}\rangle^2 {:} 
-i \left\langle \widetilde{\nu}, \widetilde{T_c} \widetilde{\nu}^*\right\rangle,
\end{multline*}
so that
\begin{multline}\label{<T(<v,phi>)>_0}
\left\langle T\left( \langle \widetilde{\nu},\widetilde{\boldsymbol{\varphi}}\rangle^2\right)\right\rangle_{{}_{0}}
=
-i \sum\limits_{\widehat{l},\widehat{n}, i,j} (2l+1) 
\widetilde{\nu}(\widehat{n}\cdot\widehat{l})_{{}_{j \, i}} \,\,
\sqrt{4\pi} \widetilde{\Delta_{c}}(\widehat{n}\cdot\widehat{l})_{{}_{jj}}
\,\,
\widetilde{\nu}(\widehat{-n}\cdot\widehat{l})_{{}_{-j \, -i}}
\\
=
-i \sum\limits_{\widehat{l},\widehat{n}} (2l+1) 
\textrm{Tr} \, \left[
\widetilde{\nu}(\widehat{n}\cdot\widehat{l}) \,\,
\widetilde{T_{c}}(\widehat{n}\cdot\widehat{l})
\,\,
\widetilde{\nu}(\widehat{n}\cdot\widehat{l})^*
\right] 
= -i \left\langle \widetilde{\nu}, \widetilde{T_c} \widetilde{\nu}^*\right\rangle.
\end{multline} 
Here ${:} \ldots {:}$ denotes the Wick ordered product, and we have introduced 
the following multiplication operator $\widetilde{T_c}$ with the following (matrix) multiplication 
components
\[
\widetilde{T_{c}}(\widehat{n}\cdot\widehat{l})_{{}_{i \, j}} = \delta_{{}_{i \, j}}
\sqrt{4\pi} \,\, \widetilde{\Delta_{c}}(\widehat{n}\cdot\widehat{l})_{{}_{jj}},
\]
and where $M^*$  denotes the Hermitian conjugation of the matrix $M$.   
Note that in the pairings 
$\langle \widetilde{\nu},\widetilde{\boldsymbol{\varphi}}\rangle, \langle \widetilde{\nu},
\widetilde{\boldsymbol{\varphi}}\rangle^2, \ldots$ involving free field operator $\widetilde{\boldsymbol{\varphi}}$,
we have the summation range of the characters $\widehat{n}\cdot\widehat{l}$ restricted to the positive energy orbit $\mathscr{O}^+$, 
in accordance with the support of the free field operator components of $\widetilde{\boldsymbol{\varphi}}$, while in the pairing 
$-i \left\langle \widetilde{\nu}, \widetilde{T_c} \widetilde{\nu}^*\right\rangle$ equal to the chronological
pairing, we have, in principle, the full summation range over \emph{all} characters $\widehat{n}\cdot\widehat{l}$ of the group $G=$ EU. 
But the expression (\ref{<T(<v,phi>)>_0}) is meaningful if $\left\langle \widetilde{\nu}, \widetilde{T_c} \widetilde{\nu}^*\right\rangle$
is convergent. This will be the case for all $\widetilde{\nu} \in E_1$ that are supported on the set $\mathscr{R}$ of characters 
$\widehat{n}\cdot\widehat{l}$ for which 
\begin{equation}\label{R}
{\textstyle\frac{1}{(n^2 +1 + l(l+1))^4}} \leq \widetilde{\Delta_{c}}(\widehat{n}\cdot\widehat{l})_{{}_{jj}} \leq (n^2 +1 + l(l+1))^2. 
\end{equation}
This is so, because for $\widetilde{\nu} \in \widetilde{E}$ supported at $\mathscr{R}$ we have the inequality
\begin{equation}\label{|<nuTcnu>|<|nu|_1^2}
\left| \left\langle \widetilde{\nu}, \widetilde{T_c} \widetilde{\nu}^*\right\rangle \right| 
\leq \, \sqrt{4\pi} \, \left| \widetilde{A} \widetilde{\nu}\right|_{{}_{0}}^{2} = \sqrt{4\pi} \, \left|\widetilde{\nu} \right|_{{}_{1}}^{2}.
\end{equation}
Here, we only use the right-hand part of the inequality (\ref{R}). The role of the left-hand inequality will be explained later.
The character range of the summation in the operator pairing  $\langle \widetilde{\nu},\widetilde{\boldsymbol{\varphi}}\rangle$
is automatically restricted to the corresponding orbit $\mathscr{O}^+$, which for the massive field on EU
is automatically a finite subset of the  set of all characters. In passing to the massless case, we generalize this
property and will assume that the allowed range $\mathscr{R}$ intersects $\mathscr{O}^+$ and $\mathscr{O}^-$ at a finite set
\begin{equation}\label{RnO+-=Finite}
\mathscr{R} \cap \mathscr{O}^+ = \mathscr{O}^{+}_{0} = \textrm{finite set},
\,\,\,\,\,     \mathscr{R} \cap \mathscr{O}^- = \mathscr{O}^{-}_{0} = \textrm{finite set}.
\end{equation}
More precisely, we assume existence
of $R>0$ such that all $\widehat{n}\cdot\widehat{l} \in \mathscr{O}^{\pm}$ with $n^2 +1 + l(l+1) < R$ 
are contained in $\mathscr{R}$ and all characters of $\mathscr{O}^{\pm}$ with  $n^2 +1 + l(l+1) > R$ are not contained in $\mathscr{R}$.
Also, it will be sufficient to fulfill the condition  (\ref{R}) asymptotically: we assume the existence of $R'>0$,
such that (\ref{R}) is fulfilled for all $\widehat{n}\cdot\widehat{l} \in \mathscr{R}$ with $n^2 +1 + l(l+1) > R'$.

In order to compute the chronological product $T\left( \langle \widetilde{\nu},\widetilde{\boldsymbol{\varphi}}\rangle^n\right) $
of the higher order monomial $\langle \widetilde{\nu},\widetilde{\boldsymbol{\varphi}}\rangle^n$, let us consider the product
\[
\widetilde{\boldsymbol{\varphi}}(\widehat{n_1}\cdot\widehat{l_1})_{{}_{i_1 \,\, j_1}} \ldots
\widetilde{\boldsymbol{\varphi}}(\widehat{n_n}\cdot\widehat{l_n})_{{}_{i_n \,\, j_n}}
\]
of $n$-order. It is well-defined as an operator transforming continuously the Hida space $(E)$ again
into the Hida space, because this is so for the Hida annihilation-creation operators on EU. 
Next, let us consider such $n$-order product containing exactly $k$ pairings. Let us denote 
the number of all $k$ pairings in this $n$-order product by $\left[\begin{smallmatrix} n \\ k \end{smallmatrix} \right]$.
We observe that
\begin{equation}\label{[n,k]}
\left[\begin{smallmatrix} n \\ k \end{smallmatrix} \right] = 
\begin{cases}
{\textstyle\frac{n!}{k!(n-2k)!2^k}}, & n \geq 2k, \\
0, & n < 2k,
\end{cases}
\end{equation}
which can be proven by induction using the following identities:
\[
\left[\begin{smallmatrix} n +1 \\ k \end{smallmatrix} \right] = \left[\begin{smallmatrix} n  \\ k \end{smallmatrix} \right]
+ n \left[\begin{smallmatrix} n -1 \\ k-1 \end{smallmatrix} \right],
\,\,\,
\left[\begin{smallmatrix} n  \\ k +1 \end{smallmatrix} \right] 
= \left[\begin{smallmatrix} n -2k  \\ 1 \end{smallmatrix} \right] {\textstyle\frac{1}{k+1}},
\,\,\,
\left[\begin{smallmatrix} n  \\ 0 \end{smallmatrix} \right] = 1, \left[\begin{smallmatrix} n  \\ 1 \end{smallmatrix} \right]
= \textstyle{\binom{n}{2}}.
\]
Again, using induction with respect to the order, and the formula (\ref{[n,k]}), we show that for each natural $n$
\begin{equation}\label{InductionCP}
\sum\limits_{k=0}^{n} T\left({\textstyle\frac{i^k\langle \widetilde{\nu},\widetilde{\boldsymbol{\varphi}}\rangle^k}{k!}}\right)
=
\textrm{all terms of} \,\, \sum\limits_{k=0}^{n} {:}{\textstyle\frac{i^k\left(\langle \widetilde{\nu},\widetilde{\boldsymbol{\varphi}}\rangle
+ \left\langle \widetilde{\nu}, \widetilde{T_c} \widetilde{\nu}^*\right\rangle/2 \right)^k}{k!}}{:} \,\,
\textrm{up to order} \, n,
\end{equation}
where each $\widetilde{\boldsymbol{\varphi}}$ is counted as a quantity of the $1$-st order and each pairing  
$\left\langle \widetilde{\nu}, \widetilde{T_c} \widetilde{\nu}^*\right\rangle/2$ is counted as a quantity of the $2$-nd order. Therefore
if the series on the r.h.s. of (\ref{InductionCP}) is convergent, then we obtain
\begin{equation}\label{T(expi<nu,phi>)}
T\left( e^{i\langle \nu,\boldsymbol{\varphi}\rangle}\right)  \,\, = \,\,\,\,  {:}e^{i\left[\langle \widetilde{\nu},\widetilde{\boldsymbol{\varphi}}\rangle
+ \left\langle \widetilde{\nu}, \widetilde{T_c} \widetilde{\nu}^*\right\rangle/2 \right]}{:} \,\,\,\, = \,\,\,\,
{:}e^{i \langle \widetilde{\nu},\widetilde{\boldsymbol{\varphi}}\rangle}{:} \,\,\,\,
e^{\frac{i}{2}\left\langle \widetilde{\nu}, \widetilde{T_c} \widetilde{\nu}^*\right\rangle}.
\end{equation}

Let us now investigate the convergence of the operator series
\begin{equation}\label{SeriesXi}
\Xi = \,\,\, {:}e^{i\left[\langle \widetilde{\nu},\widetilde{\boldsymbol{\varphi}}\rangle
+ \left\langle \widetilde{\nu}, \widetilde{T_c} \widetilde{\nu}^*\right\rangle/2 \right]}{:} \,\,\, =
\sum\limits_{k=0}^{+\infty} {:}{\textstyle\frac{i^k\left(\langle \widetilde{\nu},\widetilde{\boldsymbol{\varphi}}\rangle
+ \left\langle \widetilde{\nu}, \widetilde{T_c} \widetilde{\nu}^*\right\rangle/2 \right)^k}{k!}}{:},
\end{equation}
regarded as an operator defined on the Hida test space $(\widetilde{E})$, with the space of elements in $\widetilde{E}$ supported at the orbit $\mathscr{O}^+$ of the
free field $\boldsymbol{\varphi}$, i.e. with $\Xi$ acting in the dense subspace $(\widetilde{E})$ 
of the Fock space of the free field $\boldsymbol{\varphi}$. To this end, let us note that the operator
\[
\langle \widetilde{\nu},\widetilde{\boldsymbol{\varphi}}\rangle 
=
\sum\limits_{\widehat{n}\cdot\widehat{l} \in \mathscr{O}^+} (2l+1)
\textrm{Tr} \, \left[
\widetilde{\nu}(\widehat{n}\cdot\widehat{l})
\widetilde{\boldsymbol{\varphi}}(\widehat{n}\cdot\widehat{l}) 
\right] = 
\Xi(\kappa_{0,1}) + \Xi(\kappa_{1,0})
\]
is equal to the sum of two well-defined integral kernel operators
\[
\Xi(\kappa_{0,1}) = \sum\limits_{\widehat{n}\cdot\widehat{l} \in \mathscr{O}^+,s} \kappa_{0,1}(s,\widehat{n}\cdot\widehat{l}) \, a_{{}_{s}}(\widehat{n}\cdot\widehat{l}),
\,\,\, 
\Xi(\kappa_{1,0}) = \sum\limits_{\widehat{n}\cdot\widehat{l} \in \mathscr{O}^+,s} \kappa_{1,0}(s,\widehat{n}\cdot\widehat{l}) \, a_{{}_{s}}(\widehat{n}\cdot\widehat{l})^+
\]
acting on the Hida space $(\widetilde{E})$, again with the space of elements of $\widetilde{E}$ supported at the orbit $\mathscr{O}^+$ of the
free field $\boldsymbol{\varphi}$, i.e., in the Fock space of the free field $\boldsymbol{\varphi}$. Here
\begin{multline*}
\kappa_{0,1}(s,\widehat{n}\cdot\widehat{l}) = \sqrt{4\pi}
\sum\limits_{ -l \leq i,j \leq l} 
u_{{}_{s}}(\widehat{n}\cdot\widehat{l})_{{}_{i j}} \widetilde{\nu}(\widehat{n}\cdot\widehat{l})_{{}_{j i}},
\\
\kappa_{1,0}(s,\widehat{n}\cdot\widehat{l}) = \sqrt{4\pi}
\sum\limits_{ -l \leq i,j \leq l} 
\overline{u_{{}_{s}}(\widehat{-n}\cdot\widehat{l})_{{}_{-j \, -i}}} \widetilde{\nu}(\widehat{n}\cdot\widehat{l})_{{}_{j i}}. 
\end{multline*}
In the case of a massive scalar field, the kernels $\kappa_{0,1},\kappa_{1,0}$ 
have finite support $\mathscr{O}_0 =\{(\widehat{n}\cdot\widehat{l},s):
\widehat{n}\cdot\widehat{l} \in \mathscr{O}^+, 1 \leq s \leq (2l+1)^2\}$, and by the assumption
(\ref{RnO+-=Finite}), put on the allowable $\widetilde{\nu}$, these kernels have finite support
$\mathscr{O}_0 =\{(\widehat{n}\cdot\widehat{l},s):
\widehat{n}\cdot\widehat{l} \in \mathscr{O}^{+}_{0}, 1 \leq s \leq (2l+1)^2\}$ in general,
where $\mathscr{R} \cap\mathscr{O}^{+} = \mathscr{O}^{+}_{0}$. From this finite support property
of $\kappa_{0,1},\kappa_{1,0}$ and Theorem 2.6 of \cite{hida}, it follows that the integral kernel
operators $\Xi(\kappa_{0,1}),\Xi(\kappa_{1,0})$ transform continuously the test Hida space $(\widetilde{E})$ into itself, where
$(\widetilde{E})$ is the Hida test space $(\widetilde{E}_{{}_{\mathscr{O}^+}})$ in the Fock space of the field $\boldsymbol{\varphi}$, \emph{i.e.}
constructed upon the elements with Fourier transforms supported at $\mathscr{O}^+$.   

Recall that we have the pairings
\begin{multline*}
\langle \kappa_{0,1}, \widetilde{\xi}\rangle = \sum\limits_{\widehat{n}\cdot\widehat{l} 
\in \mathscr{O}^+,s} \kappa_{0,1}(s,\widehat{n}\cdot\widehat{l}) \, \widetilde{\xi}(s,\widehat{n}\cdot\widehat{l}),
\\
\langle \kappa_{1,0}, \widetilde{\xi}\rangle = \sum\limits_{\widehat{n}\cdot\widehat{l} 
\in \mathscr{O}^+,s} \kappa_{1,0}(s,\widehat{n}\cdot\widehat{l}) \, \widetilde{\xi}(s,\widehat{n}\cdot\widehat{l}),
\,\,\,\,
\widetilde{\xi}\in S_{\widetilde{A}}(\mathscr{O}),
\end{multline*}
and that $\widetilde{\xi}\in S_{\widetilde{A}}(\mathscr{O}) \simeq \widetilde{E}_{{}_{\mathscr{O}^+}}$ iff
\[
\widetilde{\xi}(s,\widehat{n}\cdot\widehat{l}) = U\widetilde{\phi}(s,\widehat{n}\cdot\widehat{l}) = 
(2l+1) \textrm{Tr} \, [u_{{}_{s}}(\widehat{n}\cdot\widehat{l})^*\widetilde{\phi}(\widehat{n}\cdot\widehat{l})] 
\]
for $\widetilde{\phi} \in \widetilde{E}$, supported at $\mathscr{O}^+$, with the unitary isomorphism $U$ between the standard 
realization $S_{\widetilde{A}}(\mathscr{O})$ and $\widetilde{E}_{{}_{\mathscr{O}^+}}$, induced by the matrices $u$. 
Recall that the discrete set $\mathscr{O}$ is an enlargement of the orbit
$\mathscr{O}^+$, with each character $\widehat{n}\cdot\widehat{l}\in \mathscr{O}^+$ having $(2l+1)^2$ 
copies in $\mathscr{O}$.
After identifying the isomorphic spaces, we see that
$\widetilde{\xi} \in \widetilde{E}_k$, iff 
\[
\sum\limits_{\widehat{n}\cdot\widehat{l} \in \mathscr{O}^+,s}  
\left(n^2+1 +l(l+1)\right)^{2k}\left|\widetilde{\xi}(s,\widehat{n}\cdot\widehat{l})\right|^2 < +\infty.
\] 
It is important that in each case, massive or massless field $\boldsymbol{\varphi}$, the intersection of the support 
$\mathscr{R}$ of $\widetilde{\nu}$ with the orbit $\mathscr{O}^+$, is a finite set $\mathscr{O}^{+}_{0}$ (in the massive case, this 
is automatic, because the whole $\mathscr{O}^+$ is finite itself in this case). Thus, by assumption, the support of $\kappa_{0,1}, \kappa_{1,0}$,
is in each case, finite. Therefore, the pairings 
\[
\widetilde{\xi} \longmapsto \langle \kappa_{0,1}, \widetilde{\xi}\rangle, \,\,\,\,\,
\widetilde{\xi} \longmapsto \langle \kappa_{1,0}, \widetilde{\xi}\rangle
\]
can be extended to continuous functionals on the spaces $E_{k}$ for all negative $k$,
with finite respective norms $| \cdot |_{k}$, $k \in \mathbb{Z}$, or $k \in \mathbb{R}$.

For the investigation of the convergence of the series $\Xi$ we use the symbol calculus.
Recall that the symbol of a generalized operator $\Xi$ transforming continuously $(E)$ into $(E)^*$ 
is defined in the following manner:
\[
\textrm{Symbol} \, \left[ \Xi \right] (\widetilde{\xi}, \widetilde{\eta}) = \left\langle\left\langle \,\,\, \Xi \, \epsilon_{{}_{\widetilde{\xi}}} \,\, ,  
\,\,\, \epsilon_{{}_{\widetilde{\eta}}} \,\, \right\rangle\right\rangle,
\,\,\,\,\,\, \widetilde{\xi}, \widetilde{\eta} \in \widetilde{E},
\]
where $\epsilon_{{}_{\widetilde{\eta}}} \in (\widetilde{E})_\mathbb{C}$ is the standard coherent state, \emph{i.e.}
the corresponding function on $E^*$ defining $\epsilon_{{}_{\widetilde{\eta}}}$ is equal
\[
\epsilon_{{}_{\widetilde{\eta}}}(x) = \textrm{exp} \left( \left\langle x, \widetilde{\eta}\right\rangle 
- {\textstyle\frac{1}{2}} \left\langle \widetilde{\eta}, \widetilde{\eta}\right\rangle  \right),
\,\,\,\, x \in E^*, \, \widetilde{\eta} \in E_\mathbb{C}.
\]
For example, the symbol of an integral kernel operator
\[
\Xi(\kappa_{l,m}), \,\,\, \kappa_{l,m} \in E^{*\widehat{\otimes} l}\otimes E^{*\widehat{\otimes} l} 
\]
is equal
\[
\textrm{Symbol} \, \left[ \Xi(\kappa_{l,m})\right](\widetilde{\xi}, \widetilde{\eta}) = e^{\langle \widetilde{\xi}, \widetilde{\eta}\rangle}
\left\langle\kappa_{l,m}, \,\, \widetilde{\eta}^{\otimes l} \otimes \widetilde{\xi}^{\otimes m}\right\rangle.
\]
In this place we are using the Hida spaces constructed
with the spaces of elements supported at the orbit $\mathscr{O}^+$. 

By the fact that the kernel $\kappa$ of the Wick product 
\[
{:}\Xi(\kappa_{l,m})\Xi(\kappa'_{l',m'}){:}
\]
of integral kernel operators 
\[
\Xi(\kappa_{l,m}),\Xi(\kappa'_{l',m'})
\]
is equal to the symmetrized tensor product $\kappa = \kappa_{l,m}\widehat{\otimes}\kappa'_{l',m'}$ 
of the kernels $\kappa_{l,m},\kappa'_{l',m'}$, we obtain
\begin{multline*}
e^{-\langle \widetilde{\xi}, \widetilde{\eta}\rangle} \,\, 
\textrm{Symbol} \, \left[{:}{\textstyle\frac{i^n\left(\langle \widetilde{\nu},\widetilde{\boldsymbol{\varphi}}\rangle
+ \left\langle \widetilde{\nu}, \widetilde{T_c} \widetilde{\nu}^*\right\rangle/2 \right)^n}{n!}}{:}  \right](\widetilde{\xi}, \widetilde{\eta})
\\
=
{\textstyle\frac{i^n}{n!}} \left(  
\langle \kappa_{0,1}, \widetilde{\xi}\rangle + \langle \kappa_{1,0}, \widetilde{\eta}\rangle 
+ \textstyle{\frac{1}{2}} \left\langle \widetilde{\nu}, \widetilde{T_c} \widetilde{\nu}^*\right\rangle 
\right)^n
\end{multline*}
and for the series $\Xi$, (\ref{SeriesXi}),  we obtain 
\[
\textrm{Symbol} \, \left[ \Xi \right] (\widetilde{\xi}, \widetilde{\eta}) = 
e^{\langle \widetilde{\xi}, \widetilde{\eta}\rangle} 
\exp i \left[\langle \kappa_{0,1}, \widetilde{\xi}\rangle + \langle \kappa_{1,0}, \widetilde{\eta}\rangle 
+ \textstyle{\frac{1}{2}} \left\langle \widetilde{\nu}, \widetilde{T_c} \widetilde{\nu}^*\right\rangle \right].
\]

\begin{lem}
For any $p\geq 0$ and $\varepsilon >0$, there exist $C \geq 0$ and $q \geq 0$ such that
\begin{multline*}
\left| 
\exp i\left[\langle \kappa_{0,1}, \widetilde{\xi}\rangle + \langle \kappa_{1,0}, \widetilde{\eta}\rangle 
+ \textstyle{\frac{1}{2}} \left\langle \widetilde{\nu}, \widetilde{T_c} \widetilde{\nu}^*\right\rangle \right] 
\right|
\\
\leq 
\,\,\,\,\,\,\,\,\, 
C  \exp \varepsilon \left(
\left| \widetilde{\xi} \right|_{{}_{p+q}}^{2}
+
\left| \widetilde{\eta} \right|_{{}_{-p}}^{2}
\right),
\,\,\,\,\,\,\,\,\,\,\,\,\,\,\,\,\,\,\,\,\,\,\,\,\,\,\,\,\,\,\,\,\,\,\,\,\, 
\widetilde{\xi}, \widetilde{\eta} \in \widetilde{E}_\mathbb{C}.
\end{multline*}
Here $\widetilde{E}_\mathbb{C}$ is the complexification of the space 
of elements of $\widetilde{E}_{{}_{\mathscr{O}^+}}$, supported at $\mathscr{O}^+$. 
$\widetilde{\xi}, \widetilde{\eta} \in \widetilde{E}_\mathbb{C}$
means that $\widetilde{\xi}, \widetilde{\eta}$ are representants of the elements of $\widetilde{E}_{{}_{\mathscr{O}^+}}$
in the standard realization $\mathcal{S}_{\widetilde{A}}(\mathscr{O})$ of 
$\widetilde{E}_{{}_{\mathscr{O}^+}}$.
\label{WickPartSymboInequality}
\end{lem}
\qedsymbol \,
Let $p\geq 0$ and $\varepsilon >0$.
By the assumption put on $\mathscr{R}$, the support $\mathscr{O}_{0}$ of $\kappa_{0,1}, \kappa_{1,0}$
is finite. Therefore, $\kappa_{0,1}, \kappa_{1,0}$,
regarded as linear maps on each fixed $E_k$, are finite dimensional, and supported at the elements 
with the series components, which are non-zero only on the finite subset $\mathscr{O}_{0}$.
It follows that for any $q \geq 0$, there exist $c_{0,1}(p+q),c_{1,0}(-p)\geq 0$, such that 
\[
\left| \langle \kappa_{0,1}, \widetilde{\xi}\rangle\right| \leq c_{0,1}(p+q) \left|\widetilde{\xi} \right|_{{}_{p+q}},
\,\,\,
\left| \langle \kappa_{1,0}, \widetilde{\eta}\rangle\right| \leq c_{1,0}(-p) \left| \widetilde{\eta}  \right|_{{}_{-p}},
\,\,\,
\widetilde{\eta}, \widetilde{\xi} \in E_\mathbb{C}.
\]
It is sufficient to take 
\begin{multline*}
c_{0,1}(p+q) = \underset{\mathscr{O}_0}{\textrm{sup}} 
\left|\kappa_{0,1}(s,\widehat{n}\cdot\widehat{l})(n^2+1+l(l+1))^{-p-q}\right|,
\\
c_{1,0}(-p) = \underset{\mathscr{O}_0}{\textrm{sup}} 
\left|\kappa_{1,0}(s,\widehat{n}\cdot\widehat{l})(\widehat{n}\cdot\widehat{l})(n^2+1+l(l+1))^{p}\right|.
\end{multline*}
For all $\widetilde{\eta}, \widetilde{\xi} \in E_\mathbb{C}$ such that
\begin{equation}\label{xi,eta-BoundedDomain}
{\textstyle\frac{c_{1,0}(p+q) }{\varepsilon}} \,\,\,\,\,\,\, <
\,\,\,\,\,\, \left| \widetilde{\xi}  \right|_{{}_{p+q}}, 
\,\,\,\,\,\,\,\,\,\,\,\,\,\,\,\,\,\,\,\,\,\,\,\,\,\,\,\,\,\,\,\,\,\,
{\textstyle\frac{c_{1,0}(-p) }{\varepsilon}} \,\,\,\,\,\,\,\,
< \,\,\,\,\,\,\,\, \left| \widetilde{\eta}  \right|_{{}_{-p}} 
\end{equation}
we have
\begin{multline*}
\left| \langle \kappa_{0,1}, \widetilde{\xi}\rangle\right|
+
\left| \langle \kappa_{1,0}, \widetilde{\eta}\rangle\right|
\\
\,\,\,\,\,\,\,\,\,\,\,\,\,\,\,
\leq
\,\,\,\,\,\,\,\,\,\,\,\,\,\,\,
c_{0,1}(p+q) \left|\widetilde{\xi} \right|_{{}_{p+q}}
+
c_{1,0}(-p) \left| \widetilde{\eta}  \right|_{{}_{-p}}
\,\,\,\,\,
<
\,\,\,\,\,\,\,\,\,\,\,\,\,\,\,
\varepsilon \left( 
\left| \widetilde{\xi} \right|_{{}_{p+q}}^{2}
+
\left| \widetilde{\eta} \right|_{{}_{-p}}^{2}
\right).
\end{multline*}
Therefore, for each $p\geq 0$ and $\varepsilon >0$, we can find $C, q\geq 0$, such that
the inequality of the lemma holds in the domain (\ref{xi,eta-BoundedDomain}). It is immediate
that for fixed $p,\varepsilon$ we can find appropriate $C,q$, on the bounded completion 
\[
{\textstyle\frac{c_{1,0}(p+q) }{\varepsilon}} \,\,\,\,\,\,\,
\geq \,\,\,\,\,\, \left| \widetilde{\xi}  \right|_{{}_{p+q}}, 
\,\,\,\,\,\,\,\,\,\,\,\,\,\,\,\,\,\,\,\,\,\,\,\,\,\,\,\,\,\,\,\,\,\,
{\textstyle\frac{c_{1,0}(-p) }{\varepsilon}} \,\,\,\,\,\,\
\geq \,\,\,\,\,\,\, \left| \widetilde{\eta}  \right|_{{}_{-p}} 
\]
of the domain (\ref{xi,eta-BoundedDomain}), with the function on the l.h.s. of the inequality of the lemma
being continuous on the bounded closed subset of a finite dimensional linear space. 
Choosing the maximal values of $C,q$, which fit the inequality
of lemma separately in these two domains, we construct $C,q$, required by the lemma.
\qed

\begin{lem}
Let $\widetilde{E}_\mathbb{C}$ be the complexification of the space of elements of $\widetilde{E}$, 
supported at $\mathscr{O}^+$. Let $\widetilde{\xi}, \widetilde{\eta}$ be representants of the elements of $\widetilde{E}_{{}_{\mathscr{O}^+}}$ in the
standard realization $\mathcal{S}_{\widetilde{A}}(\mathscr{O})$ of 
$\widetilde{E}_{{}_{\mathscr{O}^+}}$.  

For any $p\geq 0$ and $\varepsilon >0$, there exist $C \geq 0$ and $q \geq 0$ such that
\[
\left| 
e^{\langle \widetilde{\xi}, \widetilde{\eta}\rangle} 
\right|
\,\,\,\,\, 
\leq 
\,\,\,\,\, 
C  \exp \varepsilon \left(
\left| \widetilde{\xi} \right|_{{}_{p+q}}^{2}
+
\left| \widetilde{\eta} \right|_{{}_{-p}}^{2}
\right),
\]
for all $\widetilde{\xi}, \widetilde{\eta} \in \mathcal{S}_{\widetilde{A}}(\mathscr{O})$,
with the support contained in the finite set $\mathscr{O}_0$.
\label{ScalarPartSymboInequality}
\end{lem}
\qedsymbol \,
For each point $j$ of the discrete support set $\mathscr{O}_0$ we introduce an auxiliary integral kernel
operator $\Xi(\kappa_{0,1})$, with the kernel $\kappa_{0,1}$ supported at this point $j$ and having the value $1$
at this point. By Theorem 2.6, \cite{hida}, each such $\Xi(\kappa_{0,1})$ transforms continuously
$(\widetilde{E})$ into $(\widetilde{E})$, where we mean Hida space  $(\widetilde{E}_{{}_{\mathscr{O}^+}})$,
and use the standard realization $\mathcal{S}_{\widetilde{A}}(\mathscr{O})$ of $\widetilde{E}_{{}_{\mathscr{O}^+}}$.
Therefore, the symbol of each such auxiliary operator respects the following condition of \cite{obataJFA}, \S 4.2.
For any $p\geq 0$ and $\varepsilon >0$, there exist $C_j \geq 0$ and $q_j \geq 0$ such that (\cite{obataJFA}, \S 4.2)
\begin{multline*}
\left|
\textrm{Symbol} \, \left[ \Xi(\kappa_{0,1}) \right] (\widetilde{\xi}, \widetilde{\eta}) 
\right|
=
\left| 
e^{\langle \widetilde{\xi}, \widetilde{\eta}\rangle}
\langle \kappa_{1,0}, \widetilde{\xi}\rangle 
\right|
\\
\leq 
\,\,\,\,\,\,\,\,\, 
C_j  \exp \varepsilon \left(
\left| \widetilde{\xi} \right|_{{}_{p+q_j}}^{2}
+
\left| \widetilde{\eta} \right|_{{}_{-p}}^{2}
\right),
\,\,\,\,\,\,\,\,\,\,\,\,\,\,\,\,\,\,\,\,\,\,\,\,\,\,\,\,\,\,\,\,\,\,\,\,\, 
\widetilde{\xi}, \widetilde{\eta} \in \widetilde{E}_\mathbb{C}.
\end{multline*}
Now we put
\[
C = \underset{j \in \mathscr{O}_0}{\textrm{max}} C_j,
\,\,\,
q = \underset{j \in \mathscr{O}_0}{\textrm{max}} q_j.
\]
Thus, for any $p\geq 0$ and $\varepsilon >0$, we constructed $C \geq 0$ and $q \geq 0$ such that
\[
\left| 
e^{\langle \widetilde{\xi}, \widetilde{\eta}\rangle}
\langle \kappa_{1,0}, \widetilde{\xi}\rangle 
\right|
\,\,\,\,\,
\leq 
\,\,\,\,\,\,\,\,\, 
C  \exp \varepsilon \left(
\left| \widetilde{\xi} \right|_{{}_{p+q}}^{2}
+
\left| \widetilde{\eta} \right|_{{}_{-p}}^{2}
\right),
\]
for all $\widetilde{\xi}, \widetilde{\eta} \in \mathcal{S}_{\widetilde{A}}(\mathscr{O})$,
with the support contained in the finite set $\mathscr{O}_0$,
and for each of the auxiliary integral kernel operators. In particular, the last inequality
holds for all elements $\widetilde{\xi}, \widetilde{\eta}$ in the assertion of the lemma 
which moreover respect
\begin{equation}\label{Boundedxi}
\left| \widetilde{\xi}\right|_{{}_{0}}^{2} > \sharp \mathscr{O}_0,
\,\,\,\,
\left| \widetilde{\eta}\right|_{{}_{0}}^{2} > \sharp \mathscr{O}_0,
\end{equation}
where $\sharp \mathscr{O}_0$ is the number of elements of $\mathscr{O}_0$. But each $\widetilde{\xi}$,
supported in $\mathscr{O}_0$ and fulfilling (\ref{Boundedxi}), 
must have at least one component, say $j$-th, with the absolute value greater than one,
with $|\langle \kappa_{1,0}, \widetilde{\xi}\rangle | >1$, where $\kappa_{1,0}$ is the kernel supported at $j$.
Therefore,  for any $p\geq 0$ and $\varepsilon >0$, we thus constructed $C \geq 0$ and $q \geq 0$ such that
\[
\left| 
e^{\langle \widetilde{\xi}, \widetilde{\eta}\rangle}
\right|
\,\,\,\,\,
\leq 
\,\,\,\,\,\,\,\,\, 
C  \exp \varepsilon \left(
\left| \widetilde{\xi} \right|_{{}_{p+q}}^{2}
+
\left| \widetilde{\eta} \right|_{{}_{-p}}^{2}
\right),
\]
for all $\widetilde{\xi}, \widetilde{\eta} \in \mathcal{S}_{\widetilde{A}}(\mathscr{O})$,
with the support contained in the finite set $\mathscr{O}_0$, which moreover respect (\ref{Boundedxi}).
On the domain complementary to (\ref{Boundedxi}), the l.h.s.and the r.h.s of the inequality of lemma are
equal to absolute values of continuous functions
on a compact subset of a finite-dimensional linear space, and for each $p\geq 0$, $\varepsilon >0$, 
we can find appropriate $C,q$ for this domain. Taking again the maximal $C,q$, corresponding to these domains,
we construct $C,q$, required by the lemma.  
\qed

\begin{lem}
For any $p\geq 0$ and $\varepsilon >0$, there exist $C \geq 0$ and $q \geq 0$ such that
\begin{multline*}
\left| 
\textrm{Symbol} \, \left[ \Xi \right] (\widetilde{\xi}, \widetilde{\eta})
\right|
 =
\left| 
e^{\langle \widetilde{\xi}, \widetilde{\eta}\rangle} 
\exp i \left[\langle \kappa_{0,1}, \widetilde{\xi}\rangle + \langle \kappa_{1,0}, \widetilde{\eta}\rangle 
+ \textstyle{\frac{1}{2}} \left\langle \widetilde{\nu}, \widetilde{T_c} \widetilde{\nu}^*\right\rangle \right]
\right|
\\
\leq 
\,\,\,\,\,\,\,\,\, 
C  \exp \varepsilon \left(
\left| \widetilde{\xi} \right|_{{}_{p+q}}^{2}
+
\left| \widetilde{\eta} \right|_{{}_{-p}}^{2}
\right),
\,\,\,\,\,\,\,\,\,\,\,\,\,\,\,\,\,\,\,\,\,\,\,\,\,\,\,\,\,\,\,\,\,\,\,\,\, 
\widetilde{\xi}, \widetilde{\eta} \in \widetilde{E}_\mathbb{C}.
\end{multline*}
Here $\widetilde{E}_\mathbb{C}$ is the complexification of the space 
of elements of $\widetilde{E}_{{}_{\mathscr{O}^+}}$, supported at $\mathscr{O}^+$. 
$\widetilde{\xi}, \widetilde{\eta} \in \widetilde{E}_\mathbb{C}$
means that $\widetilde{\xi}, \widetilde{\eta}$ are representants of the elements of $\widetilde{E}_{{}_{\mathscr{O}^+}}$
in the standard realization $\mathcal{S}_{\widetilde{A}}(\mathscr{O})$ of 
$\widetilde{E}_{{}_{\mathscr{O}^+}}$.
\label{SymboInequality}
\end{lem}
\qedsymbol \,
This lemma follows from lemmas \ref{WickPartSymboInequality} and \ref{ScalarPartSymboInequality}.
\qed

\begin{twr}
Let the support $\mathscr{R}$ of $\widetilde{\nu}$ respects the assumptions (\ref{R}) and (\ref{RnO+-=Finite}).
Then the series representing the following vacuum averages of the following chronological products:
\[
\left\langle T\left( \langle \widetilde{\nu},\widetilde{\boldsymbol{\varphi}}\rangle^2\right)\right\rangle_{{}_{0}}
= -i \left\langle \widetilde{\nu}, \widetilde{T_c} \widetilde{\nu}^*\right\rangle,
\]
\[
\left\langle
T\left( e^{i\langle \nu,\boldsymbol{\varphi}\rangle}\right)
\right\rangle_{{}_{0}}  \,\, = \,\,\,\,  
e^{\frac{i}{2}\left\langle \widetilde{\nu}, \widetilde{T_c} \widetilde{\nu}^*\right\rangle},
\]
are convergent for all $\widetilde{\nu} \in \widetilde{E}_1$, supported at $\mathscr{R}$.

Let, moreover, $(\widetilde{E})$ be the Hida space in the Fock space of the free field $\boldsymbol{\varphi}$, with 
the space of elements of $\widetilde{E}$ supported at $\mathscr{O}^+$.
For each $\Phi \in (\widetilde{E})$ the series  
\[
T\left( e^{i\langle \nu,\boldsymbol{\varphi}\rangle}\right)\Phi = \,\,\, {:}e^{i\left[\langle \widetilde{\nu},\widetilde{\boldsymbol{\varphi}}\rangle
+ \left\langle \widetilde{\nu}, \widetilde{T_c} \widetilde{\nu}^*\right\rangle/2 \right]}{:} \Phi \,\,\, =
\sum\limits_{k=0}^{+\infty} {:}{\textstyle\frac{i^k\left(\langle \widetilde{\nu},\widetilde{\boldsymbol{\varphi}}\rangle
+ \left\langle \widetilde{\nu}, \widetilde{T_c} \widetilde{\nu}^*\right\rangle/2 \right)^k}{k!}}{:}\Phi
\]
converges in $(\widetilde{E})$, and the operator 
\[
T\left( e^{i\langle \nu,\boldsymbol{\varphi}\rangle}\right)
\]
maps continuously the Hida space $(\widetilde{E})$ into 
$(\widetilde{E})$. 
\label{TProdConvergence}
\end{twr}
\qedsymbol \, 
The proof of the part concerning vacuum averages we have already given above. The proof concerning convergence
of the chronological product follows from theorem 4.4 of \cite{obataJFA}, and the property of the symbol
of the operator
\[
\Xi = T\left( e^{i\langle \nu,\boldsymbol{\varphi}\rangle}\right),
\]
given in lemma \ref{SymboInequality}.
\qed

Before we pass to the functional integral form of (\ref{T(expi<nu,phi>)}), let us only remark that 
the guessing method of the validity of the formulas of theorem \ref{TProdConvergence}, 
which is given in \cite{Bogoliubov_Shirkov}, \S 43.2, works pretty well also on the compactified EU.
We skip its presentation, as its adaptation to EU is quite obvious.

\section{Rigorous construction of the Feynman integral}

Now we pass to the construction of the functional integral form of (\ref{T(expi<nu,phi>)}), following
the idea of \cite{Bogoliubov_Shirkov}, \S 43.3. Using reality and parity of $\Delta_c$,
 (\ref{T(expi<nu,phi>)}) can be written in the following form:
\begin{multline}\label{Product}
\left\langle
T\left( e^{i\langle \widetilde{\nu},\widetilde{\boldsymbol{\varphi}}\rangle}\right)
\right\rangle_{{}_{0}}  \,\, = \,\,\,\,  
e^{\frac{i}{2}\left\langle \widetilde{\nu}, \widetilde{T_c} \widetilde{\nu}^*\right\rangle}
\\
= \exp \left[
{\textstyle\frac{i}{2}}
\sum\limits_{\widehat{l},\widehat{n}, i,j} (2l+1) 
\widetilde{\nu}(\widehat{n}\cdot\widehat{l})_{{}_{i \, j}} \,\,
\sqrt{4\pi} \widetilde{\Delta_{c}}(\widehat{n}\cdot\widehat{l})_{{}_{jj}}
\,\,
\widetilde{\nu}(\widehat{-n}\cdot\widehat{l})_{{}_{-i \, -j}}
\right]
\\
=
\underset{\widehat{l},\widehat{n}, i,j}{\prod} \exp 
\left[ 
{\textstyle\frac{i}{2}}
(2l+1) 
\widetilde{\nu}(\widehat{n}\cdot\widehat{l})_{{}_{i \, j}} \,\,
\sqrt{4\pi} \widetilde{\Delta_{c}}(\widehat{n}\cdot\widehat{l})_{{}_{jj}}
\,\,
\widetilde{\nu}(\widehat{-n}\cdot\widehat{l})_{{}_{-i \, -j}}
\right]
\\
=
\underset{\widehat{l},\widehat{n \geq 0}, i,j}{\prod} \exp 
\left[ 
i
(2l+1) 
\widetilde{\nu}(\widehat{n}\cdot\widehat{l})_{{}_{i \, j}} \,\,
\sqrt{4\pi} \widetilde{\Delta_{c}}(\widehat{n}\cdot\widehat{l})_{{}_{jj}}
\,\,
\widetilde{\nu}(\widehat{-n}\cdot\widehat{l})_{{}_{-i \, -j}}
\right].
\end{multline}
We consider now a fixed factor of the last product, with fixed $n,l,j,i$ and express its value in terms of a two-dimensional
Gaussian-like integral over two real variables $x_{{}_{\widehat{n},\widehat{l},j,i}}, y_{{}_{\widehat{n},\widehat{l},j,i}}$,
which we interpret, after \cite{Bogoliubov_Shirkov}, as, respectively, the real and imaginary part
of the Fourier components
\[
\widetilde{\varphi}(\widehat{n}\cdot\widehat{l})_{{}_{j \, i}}
= x_{{}_{\widehat{n},\widehat{l},j,i}} + i y_{{}_{\widehat{n},\widehat{l},j,i}}
\]
of a real function $\varphi$, therefore, respecting
\[
x_{{}_{\widehat{-n},\widehat{l},-j,-i}} = x_{{}_{\widehat{n},\widehat{l},j,i}},
\,\,\,
y_{{}_{\widehat{-n},\widehat{l},-j,-i}} = -y_{{}_{\widehat{n},\widehat{l},j,i}}.
\]
To this end,  we introduce the following parameters $\lambda,\mu,a$ of the Gaussian exponent in the Gaussian-like
integral expressing our factor 
\[
\lambda = \left[
\widetilde{\nu}(\widehat{n}\cdot\widehat{l})_{{}_{i \, j}}
+
\widetilde{\nu}(\widehat{-n}\cdot\widehat{l})_{{}_{-i \, -j}}
\right](2l+1),
\]
\[
\mu = \left[
\widetilde{\nu}(\widehat{n}\cdot\widehat{l})_{{}_{i \, j}}
-
\widetilde{\nu}(\widehat{-n}\cdot\widehat{l})_{{}_{-i \, -j}}
\right](2l+1),
\,\,\,
a = {\textstyle\frac{2l+1}{\sqrt{4\pi} \widetilde{\Delta_{c}}(\widehat{n}\cdot\widehat{l})_{{}_{jj}}}},
\]
so that
\[
\lambda^2-\mu^2 = (2l+1)^2 4 \widetilde{\nu}(\widehat{n}\cdot\widehat{l})_{{}_{i \, j}}
\widetilde{\nu}(\widehat{-n}\cdot\widehat{l})_{{}_{-i \, -j}},
\]
\[
\widetilde{\varphi}(\widehat{n}\cdot\widehat{l})_{{}_{j \, i}}\widetilde{\varphi}(\widehat{-n}\cdot\widehat{l})_{{}_{-j \, -i}}
= x_{{}_{\widehat{n},\widehat{l},j,i}}^{2} + x_{{}_{\widehat{n},\widehat{l},j,i}}^{2},
\]
\[
\lambda x_{{}_{\widehat{n},\widehat{l},j,i}} 
+ i\mu y_{{}_{\widehat{n},\widehat{l},j,i}} 
=
(2l+1)\left[
\widetilde{\nu}(\widehat{n}\cdot\widehat{l})_{{}_{i \, j}}\widetilde{\varphi}(\widehat{n}\cdot\widehat{l})_{{}_{j \, i}}
+
\widetilde{\nu}(\widehat{-n}\cdot\widehat{l})_{{}_{-i \, -j}}\widetilde{\varphi}(\widehat{-n}\cdot\widehat{l})_{{}_{-j \, -i}}
\right]
\]
Using the standard Gaussian-like quadrature, we obtain
\begin{multline*}
\exp \left[ 
i
\widetilde{\nu}(\widehat{n}\cdot\widehat{l})_{{}_{i \, j}} \,\,
\sqrt{4\pi} \widetilde{\Delta_{c}}(\widehat{n}\cdot\widehat{l})_{{}_{jj}}
\,\,
\widetilde{\nu}(\widehat{-n}\cdot\widehat{l})_{{}_{-i \, -j}}
\right] = e^{i\frac{\lambda^2-\mu^2}{4a}}
\\
=
\underset{\epsilon \rightarrow 0}{\textrm{lim}}
{\textstyle\frac{ia}{\pi}}
\int\limits_{-\infty}^{+\infty} dx_{{}_{\widehat{n},\widehat{l},j,i}} \int\limits_{-\infty}^{+\infty}
dy_{{}_{\widehat{n},\widehat{l},j,i}}
\exp
i
\left[
(-2a+i\epsilon)
{\textstyle\frac{1}{2}}
\left( x_{{}_{\widehat{n},\widehat{l},j,i}}^{2} + y_{{}_{\widehat{n},\widehat{l},j,i}}^{2} \right)
+\lambda x_{{}_{\widehat{n},\widehat{l},j,i}} 
+ i\mu y_{{}_{\widehat{n},\widehat{l},j,i}} 
\right]
\end{multline*}
\begin{multline*}
=
{\textstyle\frac{i(2l+1)}{\sqrt{4\pi^{3}}\widetilde{\Delta_{c}}(\widehat{n}\cdot\widehat{l})_{{}_{jj}}}}
\int\limits_{-\infty}^{+\infty} dx_{{}_{\widehat{n},\widehat{l},j,i}} \int\limits_{-\infty}^{+\infty}
dy_{{}_{\widehat{n},\widehat{l},j,i}}
\Bigg\lbrace
\exp
{\textstyle\frac{i}{2}}
\left[
(-{\textstyle\frac{2l+1}{\sqrt{\pi} \widetilde{\Delta_{c}}(\widehat{n}\cdot\widehat{l})_{{}_{jj}}}})
\left( x_{{}_{\widehat{n},\widehat{l},j,i}}^{2} + y_{{}_{\widehat{n},\widehat{l},j,i}}^{2} \right)
\right] 
\,\, \times
\\
\times \,\,
\exp i \left[
(2l+1)\widetilde{\nu}(\widehat{n}\cdot\widehat{l})_{{}_{i \, j}}\widetilde{\varphi}(\widehat{n}\cdot\widehat{l})_{{}_{j \, i}}
+
(2l+1)\widetilde{\nu}(\widehat{-n}\cdot\widehat{l})_{{}_{-i \, -j}}\widetilde{\varphi}(\widehat{-n}\cdot\widehat{l})_{{}_{-j \, -i}}
\right]
\Bigg\rbrace
\end{multline*}

\begin{multline*}
=
{\textstyle\frac{i(2l+1)}{\sqrt{4\pi^{3}}\widetilde{\Delta_{c}}(\widehat{n}\cdot\widehat{l})_{{}_{jj}}}}
\int\limits_{-\infty}^{+\infty} dx_{{}_{\widehat{n},\widehat{l},j,i}} \int\limits_{-\infty}^{+\infty}
dy_{{}_{\widehat{n},\widehat{l},j,i}}
\Bigg\lbrace
\exp
{\textstyle\frac{i}{2}}
\left[
\widetilde{\varphi}(\widehat{n}\cdot\widehat{l})_{{}_{j \, i}}
(-{\textstyle\frac{2l+1}{\sqrt{\pi} \widetilde{\Delta_{c}}(\widehat{n}\cdot\widehat{l})_{{}_{jj}}}})
\widetilde{\varphi}(\widehat{-n}\cdot\widehat{l})_{{}_{-j \, -i}}
\right] 
\,\, \times
\\
\times \,\,
\exp i \left[
(2l+1)\widetilde{\nu}(\widehat{n}\cdot\widehat{l})_{{}_{i \, j}}\widetilde{\varphi}(\widehat{n}\cdot\widehat{l})_{{}_{j \, i}}
+
(2l+1)\widetilde{\nu}(\widehat{-n}\cdot\widehat{l})_{{}_{-i \, -j}}\widetilde{\varphi}(\widehat{-n}\cdot\widehat{l})_{{}_{-j \, -i}}
\right]
\Bigg\rbrace
\end{multline*}
where in the last two integrals, $\epsilon$ and the limit $\epsilon \rightarrow 0$, is not explicitly written. 

Let us introduce, for each $n\geq 0$, $l$, $-l \leq j,i \leq -l$, the following
two-dimensional normalized ``complex measure''
\begin{multline*}
\delta\widetilde{\varphi}(\widehat{n}\cdot\widehat{l})_{{}_{ji}}=
{\textstyle\frac{i(2l+1)}{\sqrt{4\pi^{3}}\widetilde{\Delta_{c}}(\widehat{n}\cdot\widehat{l})_{{}_{jj}}}}
\exp
{\textstyle\frac{i}{2}}
\left[
\widetilde{\varphi}(\widehat{n}\cdot\widehat{l})_{{}_{j \, i}}
(-{\textstyle\frac{2l+1}{\sqrt{\pi} \widetilde{\Delta_{c}}(\widehat{n}\cdot\widehat{l})_{{}_{jj}}}})
\widetilde{\varphi}(\widehat{-n}\cdot\widehat{l})_{{}_{-j \, -i}}
\right] dx_{{}_{\widehat{n},\widehat{l},j,i}}dy_{{}_{\widehat{n},\widehat{l},j,i}}
\\
=
{\textstyle\frac{i(2l+1)}{\sqrt{4\pi^{3}}\widetilde{\Delta_{c}}(\widehat{n}\cdot\widehat{l})_{{}_{jj}}}}
\exp
{\textstyle\frac{i}{2}}
\left[
(-{\textstyle\frac{2l+1}{\sqrt{\pi} \widetilde{\Delta_{c}}(\widehat{n}\cdot\widehat{l})_{{}_{jj}}}})
\left( x_{{}_{\widehat{n},\widehat{l},j,i}}^{2} + y_{{}_{\widehat{n},\widehat{l},j,i}}^{2} \right)
\right] dx_{{}_{\widehat{n},\widehat{l},j,i}}dy_{{}_{\widehat{n},\widehat{l},j,i}}
\\
=
{\textstyle\frac{\exp
{\textstyle\frac{i}{2}}
\left[
(-{\textstyle\frac{2l+1}{\sqrt{\pi} \widetilde{\Delta_{c}}(\widehat{n}\cdot\widehat{l})_{{}_{jj}}}})
\left( x_{{}_{\widehat{n},\widehat{l},j,i}}^{2} + y_{{}_{\widehat{n},\widehat{l},j,i}}^{2} \right)
\right] dx_{{}_{\widehat{n},\widehat{l},j,i}}dy_{{}_{\widehat{n},\widehat{l},j,i}}}
{\int\int\exp
{\textstyle\frac{i}{2}}
\left[
(-{\textstyle\frac{2l+1}{\sqrt{\pi} \widetilde{\Delta_{c}}(\widehat{n}\cdot\widehat{l})_{{}_{jj}}}})
\left( x_{{}_{\widehat{n},\widehat{l},j,i}}^{2} + y_{{}_{\widehat{n},\widehat{l},j,i}}^{2} \right)
\right] dx_{{}_{\widehat{n},\widehat{l},j,i}}dy_{{}_{\widehat{n},\widehat{l},j,i}}}}
\end{multline*}
where again, $\epsilon$ and the limit $\epsilon \rightarrow 0$ is not explicitly written.

Returning to the product (\ref{Product}) we arrive, still at the heuristic level, with the following formula
\begin{equation}\label{BSInt}
\left\langle
T\left( e^{i\langle \widetilde{\nu},\widetilde{\boldsymbol{\varphi}}\rangle}\right)
\right\rangle_{{}_{0}}  \,\, = \,\,\,\,  
e^{\frac{i}{2}\left\langle \widetilde{\nu}, \widetilde{T_c} \widetilde{\nu}^*\right\rangle}
=
\int
e^{i\langle \widetilde{\nu},\widetilde{\varphi}\rangle}
\prod_{n\geq 0, l,j,i} \delta\widetilde{\varphi}(\widehat{n}\cdot\widehat{l})_{{}_{ji}}
\end{equation}
It is therefore tempting to search for a normalized complex measure
\[
d\mu_T(\widetilde{\varphi}) = \prod_{n\geq 0, l,j,i} \delta\widetilde{\varphi}(\widehat{n}\cdot\widehat{l})_{{}_{ji}}
\]
in the form of a product measure, first on the big space $\mathbb{R}^\infty$ of all real sequences (say, of the real and imaginary parts of 
a ``Fourier series components''), hoping that it will have support on the interesting for us linear subspace, which moreover allows us to compute
the r.h.s. of (\ref{BSInt}), and respects the condition (\ref{BSInt}):
\[
e^{\frac{i}{2}\left\langle \widetilde{\nu}, \widetilde{T_c} \widetilde{\nu}^*\right\rangle}
=
\int
e^{i\langle \widetilde{\nu},\widetilde{\varphi}\rangle}
\, d \mu_T(\widetilde{\varphi}).
\]
Although this approach may seem natural at the very first sight, we abandon it. 
We do it, because the very existence of such a product ``complex measure'', already on the huge product space
$\mathbb{R}^\infty$ is a delicate problem. Already its infinite variation (if it existed at all)
introduces additional difficulties in its construction, if we would like to get the required measure as the product measure on
$\mathbb{R}^\infty$. For example, using the $\epsilon$-limit trick, we have normalized each 
finite dimensional projection of such a hypothetical 
product measure to 1, \emph{i.e.} with the projected measure of the whole finite-dimensional subspace on which we project,
equal $1$. Initially it may thus seem that this normalization, which for probability measures on $\mathbb{R}^\infty$
saves the consistency, and by Kolmogorov's theorem, the existence of the product measure on $\mathbb{R}^\infty$, is not yet
sufficient, in our case, for the existence of the product measure.  In particular, it is immediately
seen that the sequence of measures of balls (in the finite dimensional subspace on which we project) with the radii going to infinity,
is not convergent. For a $\sigma$-measure of finite variation, it should converge to $1$. Therefore, we need to check, e.g.,
if the $\sigma$-additivity holds at least in cases in which the corresponding limits of the sums (series)
of the measures are convergent, and check if the measure is independent of countable decomposition into disjoint
measurable subsets, with the series of measures of the subsets convergent. 

In order to avoid subtleties of this construction, we are searching for
a measure on $E^* \simeq \widetilde{E}^*$, or even more generally, we are searching for a Hida distribution $P \in (E)^* \simeq(\widetilde{E})$
whose Fourier transform is equal to the l.h.s. of (\ref{BSInt}), equivalently
\[
e^{\frac{i}{2}\left\langle \widetilde{\nu}, \widetilde{T_c} \widetilde{\nu}^*\right\rangle} =
\mathcal{T}P(\widetilde{\nu}) = \int\limits_{\widetilde{E}^*} 
e^{i\langle \widetilde{\nu},\widetilde{\varphi}\rangle} P(\widetilde{\varphi})
\, d \mu(\widetilde{\varphi})
=
\left\langle\left\langle e_{{}_{\widetilde{\nu}}}, P \right\rangle\right\rangle, 
\] 
possibly with the elements $\widetilde{\varphi}, \widetilde{\nu}$ appropriately supported.
Here we have the standard Gaussian probability measure $\mu$ on $E^* \simeq \widetilde{E}^*$, which
arises in the natural Wigner-It\^o-Segal isomorphism $\Gamma(H) \simeq L^2(E^*, d\mu; \mathbb{R})$,
mentioned at the beginning of Section \ref{GelfandTriples} and where we have introduced the character function
\[
\widetilde{E}^* \ni \widetilde{\varphi} \longmapsto 
e_{{}_{\widetilde{\nu}}}(\widetilde{\varphi}) = e^{i\langle \widetilde{\nu},\widetilde{\varphi}\rangle}
\] 
of the nuclear group $\widetilde{E}^*$, well-defined for all $\widetilde{\varphi} \in \widetilde{E}^*$
if $\widetilde{\nu} \in \widetilde{E}$. Note however, that the standard Gaussian measure $\mu$
is supported at $E_{-1} \simeq \widetilde{E}_{-1} \subset \widetilde{E}^*$. In particular, the support of our distribution
$P$ (if it exists at all) is contained in $\widetilde{E}_{-1}$. Below we investigate it more closely.
Therefore, the character function $e_{{}_{\widetilde{\nu}}}$ makes sense for all $\widetilde{\varphi}$ in the support of $\mu$ 
if $\widetilde{\nu} \in \widetilde{E}_{1}$. This is also the condition for the existence of the l.h.s. of (\ref{BSInt}), as we know,
if the Fourier transform support $\mathscr{R}$ of $\widetilde{\nu}$ respects the assumptions
(\ref{R}) and (\ref{RnO+-=Finite}).
Let us remind that for $\widetilde{\eta} \in \widetilde{E}_\mathbb{C}$, 
$e_{{}_{\widetilde{\eta}}} \in (\widetilde{E})_\mathbb{C}$. Therefore we have a well-defined Fourier transform \cite{obata-book}
\[
\mathcal{T}P(\widetilde{\eta}) =
\left\langle\left\langle e_{{}_{\widetilde{\eta}}}, P \right\rangle\right\rangle
=\int\limits_{\widetilde{E}^*} 
e^{i\langle \widetilde{\eta},\widetilde{\varphi}\rangle} P(\widetilde{\varphi})
\, d \mu(\widetilde{\varphi}), \,\,\,  \eta \in \widetilde{E}_\mathbb{C}, \,\,\,
\]
of any Hida distribution $P \in (E)^* \simeq(\widetilde{E})^*$. 
The integral form of the pairing $\langle\langle \cdot , \cdot \rangle\rangle$ 
can be used here, not only purely formally,
because, similarly to finite-dimensional tempered distributions, also the infinite-dimensional
distributions $P$ can be approximated by the infinite-dimensional test functions $P_n \in (\widetilde{E})$:
\[
\mathcal{T}P(\widetilde{\eta}) =
\left\langle\left\langle e_{{}_{\widetilde{\eta}}}, P \right\rangle\right\rangle
= \underset{n \rightarrow \infty}{\textrm{lim}}
\int\limits_{\widetilde{E}^*} 
e^{i\langle \widetilde{\eta},\widetilde{\varphi}\rangle} P_n(\widetilde{\varphi})
\, d \mu(\widetilde{\varphi}).
\]

To formulate our next theorem, we need to characterize the finite-dimensional restrictions
of the distribution $P$, which should coincide with the finite-dimensional Gaussian distribution
of our heuristic construction. To this end, note the following. The system of functions 
$\eta_{{}_{\widehat{n},\widehat{l},j,i}}, \zeta_{{}_{\widehat{n},\widehat{l},j,i}}$, respectively,
with the real and imaginary parts of the Fourier transform components all equal zero
except, respectively, the real component equal $x_{{}_{\widehat{n},\widehat{l},j,i}}=1$, or, respectively, except the imaginary part component
equal $y_{{}_{\widehat{n},\widehat{l},j,i}}=1$, for a fixed $n,l,j,i$, compose a complete orthogonal system,
with the square of norms equal to $\tfrac{1}{2l+1}$. Let us denote the Fourier transform of any such system of functions
by
\[
\widetilde{\xi}_{{}_{\widehat{n},\widehat{l},j,i}} = \widetilde{\eta}_{{}_{\widehat{n},\widehat{l},j,i}} 
+\overset{\cdot}{\i} \,\, \widetilde{\zeta}_{{}_{\widehat{n},\widehat{l},j,i}}.
\]

Therefore, there arise two problems. First, we need to check if the distribution $d\mu_T(\widetilde{\varphi}) = P((\widetilde{\varphi})) d\mu(\widetilde{\varphi})$
exists. Second, if it exists, then its image under the map
\begin{multline*}
\widetilde{E}^* \ni \widetilde{\varphi} \longmapsto 
\left(
\langle \widetilde{\varphi},\eta_{{}_{\widehat{n_1},\widehat{l_1},j_1,i_1}}\rangle, 
\langle \widetilde{\varphi},\overset{\cdot}{\i} \, \zeta_{{}_{\widehat{n_1},\widehat{l_1},j_1,i_1}}\rangle, \ldots,
\langle \widetilde{\varphi},\eta_{{}_{\widehat{n_k},\widehat{l_k},j_k,i_k}}\rangle, 
\langle \widetilde{\varphi},\overset{\cdot}{\i} \, \zeta_{{}_{\widehat{n_k},\widehat{l_k},j_k,i_k}}\rangle
\right)
\\
=
(x_{{}_{\widehat{n_1},\widehat{l_1},j_1,i_1}}, y_{{}_{\widehat{n_1},\widehat{l_1},j_1,i_1}}, \ldots
x_{{}_{\widehat{n_k},\widehat{l_k},j_k,i_k}}, y_{{}_{\widehat{n_k},\widehat{l_k},j_k,i_k}})
\in \mathbb{R}^{2k}
\end{multline*}
should be equal to the following distribution
\[
\prod\limits_{i=1}^{k}
{\textstyle\frac{\overset{\cdot}{\i} \, (2l_i+1)}{\sqrt{4\pi^{3}}\widetilde{\Delta_{c}}(\widehat{n_i}\cdot\widehat{l_i})_{{}_{j_ij_i}}}}
\exp
{\textstyle\frac{\overset{\cdot}{\i}}{2}}
\left[
(-{\textstyle\frac{2l+1}{\sqrt{\pi} \widetilde{\Delta_{c}}(\widehat{n_i}\cdot\widehat{l_i})_{{}_{j_ij_i}}}})
\left( x_{{}_{\widehat{n_i},\widehat{l_i},j_i,i_i}}^{2} + y_{{}_{\widehat{n_i},\widehat{l_i},j_i,i_i}}^{2} \right)
\right].
\]
We have
\begin{twr}
If $\widetilde{\nu},\widetilde{\varphi}$ are supported at $\mathscr{R}$, which respects (\ref{R}) and 
(\ref{RnO+-=Finite}), then the distribution $P$ fulfilling the condition
\begin{equation}\label{FTmuT}
e^{\frac{\overset{\cdot}{\i}}{2}\left\langle \widetilde{\nu}, \widetilde{T_c} \widetilde{\nu}^*\right\rangle} =
\mathcal{T}P(\widetilde{\nu}) = \int\limits_{\widetilde{E}^*} 
e^{\overset{\cdot}{\i} \, \langle \widetilde{\nu},\widetilde{\varphi}\rangle} P(\widetilde{\varphi})
\, d \mu(\widetilde{\varphi})
\end{equation}
exists and is uniquely determined by this condition. 

Let 
\[
\widetilde{\xi}_{{}_{\widehat{n},\widehat{l},j,i}} = \widetilde{\eta}_{{}_{\widehat{n},\widehat{l},j,i}} 
+
\overset{\cdot}{\i} \,\, 
\widetilde{\zeta}_{{}_{\widehat{n},\widehat{l},j,i}},
\]
for $\widehat{n}\cdot\widehat{l} \in \mathscr{R}$ and with real $\widetilde{\eta},\widetilde{\zeta}$-s, 
compose an orthogonal system of functions in $\widetilde{E}$, supported at $\mathscr{R}$,
which respect the orthogonality conditions
\begin{multline*}
\langle\widetilde{\eta}_{{}_{\widehat{n},\widehat{l},j,i}}, \widetilde{\eta}_{{}_{\widehat{n'},\widehat{l'},j',i'}} \rangle
= {\textstyle\frac{1}{2l+1}}\delta_{{}_{n \, n'}}\delta_{{}_{l \, l'}}\delta_{{}_{j \, j'}}\delta_{{}_{i \, i'}},
\\
\langle\widetilde{\zeta}_{{}_{\widehat{n},\widehat{l},j,i}}, \widetilde{\zeta}_{{}_{\widehat{n'},\widehat{l'},j',i'}} \rangle
= {\textstyle\frac{1}{2l+1}} \delta_{{}_{n \, n'}}\delta_{{}_{l \, l'}}\delta_{{}_{j \, j'}}\delta_{{}_{i \, i'}},
\,\,\,
\langle\widetilde{\eta}_{{}_{\widehat{n},\widehat{l},j,i}}, \widetilde{\zeta}_{{}_{\widehat{n'},\widehat{l'},j',i'}} \rangle = 0.
\end{multline*}
Then the image of the distribution $d\mu_T(\widetilde{\varphi}) = P(\widetilde{\varphi}) d\mu(\widetilde{\varphi})$ under the map
\begin{multline*}
\widetilde{E}^* \ni \widetilde{\varphi} \longmapsto 
\left(
\langle \widetilde{\varphi},\eta_{{}_{\widehat{n_1},\widehat{l_1},j_1,i_1}}\rangle, 
\langle \widetilde{\varphi},\overset{\cdot}{\i} \, \zeta_{{}_{\widehat{n_1},\widehat{l_1},j_1,i_1}}\rangle, \ldots,
\langle \widetilde{\varphi},\eta_{{}_{\widehat{n_k},\widehat{l_k},j_k,i_k}}\rangle, 
\langle \widetilde{\varphi},\overset{\cdot}{\i} \, \zeta_{{}_{\widehat{n_k},\widehat{l_k},j_k,i_k}}\rangle
\right)
\\
=
(x_{{}_{\widehat{n_1},\widehat{l_1},j_1,i_1}}, y_{{}_{\widehat{n_1},\widehat{l_1},j_1,i_1}}, \ldots
x_{{}_{\widehat{n_k},\widehat{l_k},j_k,i_k}}, y_{{}_{\widehat{n_k},\widehat{l_k},j_k,i_k}})
\in \mathbb{R}^{2k}
\end{multline*}
is equal to the following distribution
\[
\prod\limits_{i=1}^{k}
\left[
{\textstyle\frac{\overset{\cdot}{\i} \, (2l_i+1)}{\sqrt{4\pi^{3}}\widetilde{\Delta_{c}}(\widehat{n_i}\cdot\widehat{l_i})_{{}_{j_ij_i}}}}
\exp
{\textstyle\frac{\overset{\cdot}{\i}}{2}}
\left[
(-{\textstyle\frac{2l+1}{\sqrt{\pi} \widetilde{\Delta_{c}}(\widehat{n_i}\cdot\widehat{l_i})_{{}_{j_ij_i}}}})
\left( x_{{}_{\widehat{n_i},\widehat{l_i},j_i,i_i}}^{2} + y_{{}_{\widehat{n_i},\widehat{l_i},j_i,i_i}}^{2} \right)
\right]
dx_{{}_{\widehat{n_i},\widehat{l_i},j_i,i_i}} y_{{}_{\widehat{n_i},\widehat{l_i},j_i,i_i}}
\right].
\]
\label{muTexistence}
\end{twr}

\qedsymbol \,
From the inequality (\ref{|<nuTcnu>|<|nu|_1^2}) we get the following inequalities
\[
\left| e^{ \frac{\overset{\cdot}{\i}}{2}\left\langle \widetilde{\nu}, \widetilde{T_c} \widetilde{\nu}^*\right\rangle} \right|
\leq 
e^{\frac{1}{2}\left| \left\langle \widetilde{\nu}, \widetilde{T_c} \widetilde{\nu}^*\right\rangle \right|}
\leq
e^{ \sqrt{\pi} \, \left|\widetilde{\nu} \right|_{{}_{1}}^{2}},
\]
for all complex valued $\widetilde{\nu} \in \widetilde{E}_\mathbb{C}$, supported at $\mathscr{R}$. From this and theorem 5.6.12 of \cite{obata-book}
we obtain the existence of $P$. Uniqueness follows from the fact that the kernel of the Fourier transform is zero,
\emph{i.e.} the Fourier transform $\mathcal{T}$ is injective, by Corollary 5.6.11 of \cite{obata-book}.

Let us, for simplicity of notation, introduce the following abbreviation 
\[
\widetilde{\xi}_{{}_{i}} = 
\widetilde{\xi}_{{}_{\widehat{n_i},\widehat{l_i},j_i,i_i}} = \widetilde{\eta}_{{}_{\widehat{n_i},\widehat{l_i},j_i,i_i}} 
+\overset{\cdot}{\i} \,\, \widetilde{\zeta}_{{}_{\widehat{n_i},\widehat{l_i},j_i,i_i}} = \widetilde{\eta}_{{}_{i}} 
+ \, \overset{\cdot}{\i} \, \widetilde{\zeta}_{{}_{i}}  
\]
\[
x_{{}_{i}} = x_{{}_{\widehat{n_i},\widehat{l_i},j_i,i_i}}, 
\,\,\,\,
y_{{}_{i}} = y_{{}_{\widehat{n_i},\widehat{l_i},j_i,i_i}}.
\]
Let
\[
\upsilon(x_{{}_{1}}, y_{{}_{1}}, \ldots
x_{{}_{k}}, y_{{}_{k}})
\]
be the tempered distribution on $\mathbb{R}^{2k}$, which is the image
of the distribution $P$. The finite-dimensional tempered
distribution $\upsilon$ is uniquely determined by its finite-dimensional Fourier transform $\widehat{\upsilon}$
on $\mathbb{R}^{2k}$.
Let us compute $\widehat{\upsilon}$ first and then recover $\upsilon$ by the inverse Fourier transform.
By definition, we have
\begin{multline*}
\widehat{\upsilon}(u_{{}_{1}}, w_{{}_{1}}, \ldots, u_{{}_{k}}, w_{{}_{k}}) 
\\
= 
\int\limits_{-\infty}^{+\infty} \ldots \int\limits_{-\infty}^{+\infty} \exp i \left[
\sum\limits_{i=1}^{k} u_i x_i +  w_i y_i 
\right]
\,
\upsilon(x_{{}_{1}},y_{{}_{1}}, \ldots, x_{{}_{k}}, y_{{}_{k}}) \, dx_{{}_{1}} dy_{{}_{1}} \ldots dx_{{}_{k}}dy_{{}_{k}}
\\
=
\int\limits_{\widetilde{E}^*} 
\exp \overset{\cdot}{\i}
\sum\limits_{i=1}^{k} 
\left[
\langle \widetilde{\varphi},u_i \widetilde{\eta}_{{}_{i}} \rangle + 
\langle \widetilde{\varphi},w_i \, \overset{\cdot}{\i} \, \widetilde{\zeta}_{{}_{i}} \rangle
\right]
P(\widetilde{\varphi})
\, d\mu(\widetilde{\varphi})
\\
=
\int\limits_{\widetilde{E}^*} 
\exp \overset{\cdot}{\i} \, \left\langle \widetilde{\varphi}, \sum\limits_{i=1}^{k} u_i \widetilde{\eta}_{{}_{i}} +\sum\limits_{i=1}^{k} w_i \, \overset{\cdot}{\i} \, \widetilde{\zeta}_{{}_{i}} \right\rangle
P(\widetilde{\varphi})
\, d\mu(\widetilde{\varphi}),
\end{multline*}
which by (\ref{FTmuT}) is equal to 
\begin{equation}\label{FTPojmuT}
\exp \textstyle{\frac{\overset{\cdot}{\i}}{2}}
\left[
\left\langle \sum\limits_{i=1}^{k} u_i \widetilde{\eta}_{{}_{i}} +\sum\limits_{i=1}^{k} w_i \, \overset{\cdot}{\i} \, \widetilde{\zeta}_{{}_{i}}, \widetilde{T_c} \left(\sum\limits_{i=1}^{k} u_i \widetilde{\eta}_{{}_{i}} +\sum\limits_{i=1}^{k} w_i \, \overset{\cdot}{\i} \, \widetilde{\zeta}_{{}_{i}}\right)^*\right\rangle
\right],
\end{equation}
and by orthogonality is equal
\[
\prod\limits_{i=1}^{k}
\exp \textstyle{\frac{\overset{\cdot}{\i}}{2}}
\left[
\textstyle{\frac{\sqrt{4\pi}\widetilde{\Delta_{c}}(\widehat{n_i}\cdot\widehat{l_i})_{{}_{j_ij_i}}}{2l_i+1}}
\left(u_{{}_{i}}^2 + w_{{}_{i}}^2\right)
\right]
= \widehat{\upsilon}(u_{{}_{1}}, w_{{}_{1}}, \ldots, u_{{}_{k}}, w_{{}_{k}}) 
\]
with the inverse Fourier transform 
\[
\upsilon(x_{{}_{1}},y_{{}_{1}}, \ldots, x_{{}_{k}},y_{{}_{k}})
\] 
equal to the asserted by theorem.
\qed      

Theorems \ref{TProdConvergence} and \ref{muTexistence} give a complete analysis of the
first step (\ref{BSstep1}) -- definition and existence of $\mu_T$. We pass now to the
second step (\ref{BSstep2}) -- investigation which physically interesting functionals $F$
are equal to the Fourier transform of the Hida distributions.
We note first that we have a complete
characterization of the class of functionals $F$ which
are images of the Hida distributions $\Lambda \in (\widetilde{E})^* \simeq (E)^*$. 
It is given in theorem 5.6.12 of \cite{obata-book}. Roughly speaking, it follows from this classification,
that the functional of the type
\[
F(\varphi) = e^{i\int \mathcal{L}(\varphi(x)) dx}
\]
with interaction Lagrangian density $\mathcal{L}(x)$, is in the Fourier transform image of a Hida distribution
if $\mathcal{L}$ contains natural powers of the field $\varphi(x)$ and its derivatives,
which are of second order at most in $\varphi$. This may seem to be a quite narrow class
from the physical point of view. But in the system of several interacting
fields, we can perform averaging with respect to each kind of field separately. Therefore, if each
interaction monomial in $\mathcal{L}(x)$ contains one and the same field in degree not greater than two,
then we can reach the functional as a Fourier transform, separately at each step corresponding 
to the averaging with respect to each kind
of field separately (a generalization of the Fubini theorem). In this case, we can execute the second and third steps (\ref{BSstep2}),
(\ref{BSstep3}), of the idea of \cite{Bogoliubov_Shirkov}, without essential modifications. 
In particular for the Weinberg-Salam model of electroweak interactions, exponentiation of the only term of  
$\mathcal{L}(x)$, which cannot be reached as a Fourier transform
comes from the Higgs field $\varphi$ and is of the following third order form
\begin{equation}\label{e^intphi^3}
F(\varphi) = e^{i \int \varphi(x)^3 dx}.
\end{equation}

In order to treat functionals (\ref{e^intphi^3}) of third or fourth order in $\varphi$ in the exponent, we need
to have the support of the distribution $P$ appropriately narrow, and need to know the distribution
$P$ more explicitly. By theorem \ref{muTexistence} and on heuristic grounds 
we expect $P$ to have the function-like, Gaussian-type form close to
\[
P(\widetilde{\varphi}) = e^{i \mathscr{A}_{{}_{0}}( \widetilde{\varphi})}  = e^{i \mathscr{A}_{{}_{0}}( \varphi)} 
\]   
up to a Gaussian factor ``flattening'' the standard probability measure $\mu$,
where $\mathscr{A}_{{}_{0}}$ is the action of the free classical (here scalar real) field $\varphi$, 
and which in the momentum representation
should have the form, up to normalization,  close to
\[
e^{-\frac{\overset{\cdot}{\i}}{2}\left\langle \widetilde{\varphi}, \widetilde{T_c}^{-1} \widetilde{\varphi}^*\right\rangle},
\]
up to a Gaussian factor ``flattening'' $\mu$.
For the nonrelativistic systems with one degree of freedom (and potentials equal to Fourier transforms of measures)
a similar rigorous construction of the Feynman integral has been achieved, which is based on the
standard Gaussian probability measure $\mu$. In case of potentials for which propagators are \emph{a priori} known,
it gives the formulas for the propagators, which agree with the standard formulas, \cite{Hida}. 
For such systems,  with the trajectories $y(t)$ replacing our $\varphi(x)$,  $P(y)$
has the Gaussian-type form $e^{i \mathscr{A}_{{}_{0}}(y)}$, but with the additional Brownian factor 
of Gaussian-type, ``flattening'' the probability measure $\mu$, \cite{Hida}, theorem 10.2. We expect similar
``flattening'' factor also in our case.

For this to be possible at all, the support of $P$ should be appropriately narrow, and concentrated on 
functions subjectable to a unique and natural extension of differentiation up to the second order, 
as we do have a second order differential operator in $\mathscr{A}_{{}_{0}}$ acting on $\varphi$. 
$\textrm{supp} \, P \subset \widetilde{E}_1 \simeq E_1$
would be sufficient, \emph{i.e.} $\textrm{Domain} \, A \simeq \textrm{Domain} \, \widetilde{A}$
of the self-adjoint extension of the standard operator $A$ introduced in Section \ref{GelfandTriples}. 
That the support of $P$ must be concentrated on some proper subspace $E_{k} \simeq \widetilde{E}_{k} \subset \widetilde{E}_{-1}$, 
with $k\geq 1$, can also be seen in the following way. Except for the functionals (\ref{e^intphi^3}) we need to have also
\begin{equation}\label{phi(x)phi(y)e^intphi^3}
F(\varphi) =  \varphi(x)\varphi(y)e^{i \int \varphi(z)^3 dz},
\end{equation}
for example, when computing the Green function of the Higgs boson. But the valuation
functional $\varphi \mapsto \varphi(x)$, cannot sensibly be extended (with the uniqueness of the extension achieved 
by the preservation of continuity) on the whole $E_{-1}$, but on $E_{1}$, or $E_{k}$, $k\geq 1$. 
Otherwise, we would have (\ref{phi(x)phi(y)e^intphi^3}) which is not defined on the whole
support $E_{-1}$ of $\mu$, and could not sensibly be integrated. Up to now, we have used only the right-hand inequality
of (\ref{R}). The left-hand inequality of (\ref{R}) influences the support of $P$, and we are going to exploit it now.

\begin{twr}
If the support $\mathscr{R}$ of $\widetilde{\nu}, \widetilde{\varphi}$ in (\ref{FTmuT})
 respects (\ref{R}) then $\textrm{supp} \, P \subset \widetilde{E}_1 \simeq E_1$
\end{twr}
\qedsymbol \,
From the left-hand inequality of (\ref{R}) it follows the inequality
\begin{equation}\label{TPcontinuity}
\sqrt{4\pi} \, \left|\widetilde{\nu} \right|_{{}_{-2}}^{2}
=
\sqrt{4\pi} \, \left| \widetilde{A}^2 \widetilde{\nu}\right|_{{}_{-2}}^{2} 
\leq
\left| \left\langle \widetilde{\nu}, \widetilde{T_c} \widetilde{\nu}^*\right\rangle \right|.
\end{equation}
It follows that the Fourier transform
\[
\mathcal{T} P (\widetilde{\nu}) = e^{\frac{\overset{\cdot}{\i}}{2}\left\langle \widetilde{\nu}, \widetilde{T_c} \widetilde{\nu}^*\right\rangle} 
\]
of $P$ has a continuous extension on $\widetilde{E}_{-2} \simeq E_{-2}$. Now, we observe that
the proof of Minlos' theorem 17.1 (or thm. 1.5.3 of \cite{obata-book}), as presented in \cite{Yamasaki} , 
can be extended from probability measures on the Hida distribution $P(\widetilde{\varphi}) d\mu(\widetilde{\varphi})$, 
with the finite dimensional projections equal to the standard complex Gaussian distribution of theorem \ref{muTexistence}. 
Using this generalization of Minlos' theorem and the continuous extendaibility of $\mathcal{T} P$ on 
$\widetilde{E}_{-2} \simeq E_{-2}$,  we conclude that $\textrm{supp} \, P \subset \widetilde{E}_1 \simeq E_1$,
because $A^{-1} \simeq \widetilde{A}^{-1}$ is Hilbert-Schmidt, and the inclusion $\widetilde{E}_{-1} \rightarrow \widetilde{E}_{-2}$
is Hilbert-Schmidt with the dual $E_{-1}^{*}$ of the Hilbert space $E_{-1}$ equal $E_{1}$. 

\qed

Theorem 1.5.3 is formulated in \cite{obata-book} only for probability measures on $E^* \simeq \widetilde{E}^*$. 
It is immediately seen to have a natural extension, at least on a wide class of Hida distributions. 
It is the case e.g. for distribution representable by a test Hida function, or more generally, 
by an absolutely summable function on $\widetilde{E}^*$, because the pairing is given by integration with the standard
Gaussian probability measure $\mu$. 

To treat the functional (\ref{e^intphi^3}) or (\ref{phi(x)phi(y)e^intphi^3}) we expand them, using the exponent series
into the functionals
\[
F_{n}(\widetilde{\varphi})
= \sum\limits_{k}^{n} {\textstyle\frac{{\overset{\cdot}{\i}}^k}{k!}}
\left(
\int \varphi(z)^3 dz
\right)^k
\,\,\,
\textrm{or}
\,\,\,
F_{n}(\widetilde{\varphi})
= \varphi(x)\varphi(y)\sum\limits_{k}^{n} {\textstyle\frac{{\overset{\cdot}{\i}}^k}{k!}}
\left(
\int \varphi(z)^3 dz
\right)^k
\]
which, by theorem 5.6.12 of \cite{obata-book}, are Fourier transforms of Hida distributions $\Lambda_n$:
\[
\mathcal{T}\Lambda_n(\widetilde{\varphi}) = F_n(\widetilde{\varphi}),
\]  
with $F_{n}$ converging pointwisely to $F$. Here we interrupt our analysis of convergence implicitly 
used in (\ref{BSstep3}). The idea is that the Hida distribution
calculus is largely reduced to integration techniques with respect to the standard measure $\mu$. These integrals
may be approximated by ordinary integrals with high multiplicity, on using density and orthogonality 
of the (Wick ordered) polynomials on the Gaussian space $E^* \simeq \widetilde{E}^*$.

This opens us to a method that, from the point of view of practical computations, 
may constitute an improvement of the method based on replacing space-time by a periodic finite lattice
\cite{Creutz}. As is known, the main problems associated with the method based on finite 
periodic lattices are related to the so-called artifacts. Checking whether the choice of 
UV- and IR-renormalization, given by the mere restriction to a finite periodic lattice, 
corresponds in the limit to ordinary renormalization, is problematic. 
The second main problem is related to the ambiguity in averaging over 
Fermionic variables.

Therefore, to make our solution of the problem posed in \S 43.3, \cite{Bogoliubov_Shirkov} 
interesting, we should have to give an analogous construction of $d\mu_T$ for averaging 
over Fermi fields. This can be done by extending our method, making the analog idea for averaging over Fermi fields, given 
in \S 43.4, \cite{Bogoliubov_Shirkov}, rigorous. This problem we have, to a large extent, already solved, because (following
Berezin \cite{Berezin}) we have given the analog of the Gelfand space-time triple $E \subset H \subset E^*$ of Section \ref{GelfandTriples}
with the nuclear space $E$ of Grassmann valued space-time test functions, and its nuclear dual $E^*$ of Grassmann 
valued Hida distributions, compare \cite{wawrzyckiInfinite}. 
We need to extend the isomorphism of the Wiener-It\^o-Segal from the Bose over the Fermi Fock space. Concerning
the Wick ordered polynomial functions on $E^*$ of the Bose case, 
we replace in them the pointwise (symmetric) products of ordinary 
functions with the Grassmann products of Grassmann valued functions. The lacking part is the 
analog of the standard cylindrical measure $\mu$ on $E^*$. But such  analog can be constructed 
largely along the standard lines, given in \cite{GelfandIV}, using the Berezin integration over Grassmann variables and the
transformation properties of the Gaussian integral over Grassmann variables \cite{Berezin}, analogous to the 
transformation properties of the Gaussian integral over ordinary commuting variables.

Recall that the fact that an assumption of the type (\ref{R}) should be 
imposed on the domain of the functional averaging has been clear 
to physicists from the very beginning of the application of the 
averaging method to the calculation of Green's functions,
compare \S 2.5 (\S 2.6 of Russ. Ed.) of \cite{SlavnovFaddeev}.
Possibly the condition may be to some 
extent relaxed (e.g., with slightly different power 
of the denominator on the left), but the above analysis shows that
condition of the type (\ref{R}) goes in the right direction.



\vspace*{1cm}

{\small Conflict of Interest: The authors declare that they have no
conflicts of interest.}

\end{document}